\documentclass[aps,prb,twocolumn,superscriptaddressgroupaddress,floatfix]{revtex4-2}
\usepackage{amssymb}
\usepackage{graphicx}
\usepackage{times}
\usepackage{amsmath}
\usepackage{amsthm}
\usepackage{amsfonts}
\usepackage[T1]{fontenc}
\usepackage[latin9]{inputenc}
\usepackage{array}
\usepackage{multirow}
\usepackage{color}
\usepackage{bm}
\usepackage{color}

\usepackage{bbm}
\usepackage{hyperref}
\usepackage{babel}
\usepackage[mathscr]{euscript}
\usepackage{float}
\usepackage{hyperref}
\hypersetup
{	colorlinks,%
	citecolor=green,%
	linkcolor=red,%
	urlcolor=blue%
}
\allowdisplaybreaks[4]

\begin{document}
\title{Conventional and practical metallic superconductivity arising from repulsive Coulomb coupling}
\author{Sankar Das Sarma}
\author{Jay D.\ Sau}
\author{Yi-Ting Tu}
\author{Shuyang Wang}
\affiliation{Condensed Matter Theory Center and Joint Quantum Institute, Department of Physics, University of Maryland, College Park, Maryland 20742, USA}
\begin{abstract}
A concrete question is discussed: Can there be conventional $s$-wave superconductivity in regular 3D (or 2D) metals, i.e., electrons in a jellium background, interacting via the standard Coulomb coupling?  We are interested in `practical' superconductivity that can in principle be observed in experiments, so the $T=0$ ground state being superconducting is not of interest, or for that matter a $T_c$ which is exponentially small and therefore `impractical' is also not of interest in the current work. We discuss both 2D and 3D cases, focusing mostly on the 3D case. We find that almost any theory based on the BCS-Migdal-Eliashberg paradigm, with some form of screened Coulomb coupling replacing the electron-phonon coupling in the BCS or Eliashberg theory, would uncritically predict absurdly high $T_c\sim100$~K for $s$-wave superconductivity in all metals (including the alkali metals, which are well-described by the jellium model) arising from the unavoidable fact that the Fermi, plasmon, and Coulomb potential energy scales are all $>10^4$~K. Therefore, we conclude, based on reduction ad absurdum, that the violation of the venerable Migdal theorem in this problem is sufficiently disruptive that no significance can be attached to numerous existing theoretical publications in the literature claiming plasmon-induced (or other similar Coulomb coupling-induced) practical SC. Using a careful analysis of the Eliashberg gap equations we find that the superconducting $T_c$ of the 3D (or 2D) electron gas can be reduced well below $\sim1$~K depending on choices of frequency cut-off parameters that are introduced to satisfy Migdall's theorem but are apriori unknown. The only believable result is the one discovered sixty years ago~\cite{Kohn1965} by Kohn and Luttinger predicting non-$s$-wave SC arising from Friedel oscillations with exponentially (and unobservably) low $T_c$.  We provide several theoretical approaches using both BCS and Eliashberg theories and different screening models to make our point.

\end{abstract}

\maketitle

\section{Introduction}\label{sec:intro}
Superconductivity is a ubiquitous phenomenon which has remained central to physics for more than 100 years ever since its discovery in 1911~\cite{Onnes1911}.  There are currently roughly $\sim 15{,}000$ known superconductors with $T_c$ ranging from $\sim 10$~mK to $\sim 250$~K (under high pressure).  The search for new superconducting materials with practical use is an active research area in materials science with many promises of quantum computers and large language model based artificial intelligence revolutionizing such efforts often hyped in the media.
Our use of the term 'practical' here is qualitative, implying an $s-$wave  $T_c$ which is not extremely small ( $>1K$ for example)-- when $T_c$ is very low ($\ll 1K$), any quantitative statement becomes problematic, and the emergent superconductivity is not 'practical' in a colloquial manner, ie., superconductivity which can be experimentally studied without using a dilution fridge. On a fundamental level, superconductivity, being a quintessential example of spontaneous symmetry breaking (namely, the $U(1)$ symmetry), continues to attract great attention from theorists and experimentalists alike with the ongoing recent focus being room temperature superconductors, topological superconductors, and exotic superconductors with unusual order parameter symmetries.  It may not be an exaggeration to say that superconductivity may very well be the most actively studied single topic in all of physics over the last 100 or so years.

In this context, one issue that has fascinated physicists a great deal over the years is the superconducting mechanism---what leads to the pairing glue causing the superconducting instability?  The minimal physical picture for superconductivity, and this is the relevant physics for the current work too, is that electrons pair up (``Cooper pairs''~\cite{Cooper1956}) because of an attractive interaction coupling them at the Fermi level causing the exchange of a virtual boson which acts as the glue to create the Cooper pairs.  The glued Cooper pairs then condense into a zero resistance collective ground state producing the breaking of $U(1)$ symmetry since the Cooper pairs form the condensate effectively breaking the gauge invariance.  The key physics is, however, the pairing arising from an effective attractive interaction between opposite spin electrons at the Fermi surface.  Thus, a mechanism is needed to cause an effective attractive interaction between electrons near the Fermi surface.  For most superconductors (if not all) this pairing arises from the electron-phonon interaction, which leads to a phonon-mediated effective attraction between the electron pairs with opposite spins near the Fermi surface. Since the electron spins are opposite, the spin part of the paired wavefunction is antisymmetric, making the spatial symmetry to be symmetric, and hence a singlet $s$-wave superconductivity.  All superconducting metals are such spin-singlet spatially symmetric $s$-wave superconductors, with phonons being the bosonic glue between the electron pairs.  In fact, one could probably claim that there is no known superconductor which has been compellingly shown to have a pairing glue other than electron-phonon coupling although there certainly are candidate materials where the situation remains an open question such as high-$T_c$ cuprate superconductors, heavy fermion superconductors, iron based pnictide superconductors, and perhaps various superconductors arising in multilayer graphene-type systems.

The purpose of the current work, which is partly a perspective, is to investigate critically the extent to which the electron-electron interaction by itself could lead to any `practical' $s$-wave superconductivity without the presence of any electron-phonon interaction (or other bosonic coupling outside the scope of just the electron system itself) in the Hamiltonian.  At first, the question seems absurd since the basic electron-electron interaction, i.e., Coulomb coupling, is by definition repulsive, seemingly ruling out prima facie any attractive glue for pairing, let alone the formation of a superconducting condensate of the pairs.  It turns out, however, that the situation is subtle since electrons screen each other, and the actual interaction between electrons is not the bare Coulomb interaction, but the many-body screened interaction, whose structure in general is extremely complicated.  A seminal paper~\cite{Kohn1965} by Kohn and Luttinger (KL SC) pointed out a long time ago that the ground state of an interacting electron liquid, at least for a weak short-range screened interaction, is superconducting, hinting at the possibility that interacting electron liquids may very well manifest superconductivity arising just from the Coulomb coupling.  But the KL SC is neither the conventional $s$-wave SC nor `practical' by any stretch of imagination, as the estimated $T_c$ is exponentially low for real metals ($T_c \sim 10^{-10}$ to $10^{-100}$~K), and occurs only for higher (i.e.\ not $s=0$) orbital angular momenta.  Thus, KL SC is only of academic interest, establishing a matter of principle that indeed repulsive Coulomb coupling under some circumstances can lead to a superconducting ground state arising from the existence of a sharp Fermi surface, albeit a very fragile unconventional one of no practical interest.  KL SC arises from the Friedel oscillations in screening associated with the existence of the Fermi surface which leads to $2k_F$-periodic oscillatory behavior in the statically screened Coulomb interaction, and as such can only occur not for the spherically symmetric $s$-wave SC, but for higher angular momentum orbital symmetry such $p$, $d$ wave SC.  There has been renewed interest in KL SC in the context of lattice systems such as the Hubbard model~\cite{Raghu2010} and 2D moir\'e materials~\cite{Chou2025} where the existence of van Hove singularities may enhance the KL SC $T_c$, but it is still low and the SC is always in higher orbital angular momentum channel.
We mention as an aside, however, an important point not emphasized often in the context of KL SC in 2D moir\'e materials: the Kohn-Luttinger mechanism is likely not operational in a flat band (with an effective mass of infinity) since there is no sharp Fermi surface and no Friedel oscillations in a strict flat band.

So, the KL SC, while establishing a matter of principle in the existence of SC via a strictly non-phonon mechanism, does not address the question we pose in this work:  Can there be conventional $s$-wave metallic superconductivity which is also practical (i.e.\ experimentally observable with a reasonable $T_c$)?  Much work has been done since the 1980s in the possibility of the so-called `plasmon-mediated' metallic superconductivity in electron liquids, with claims of high $T_c$ $s$-wave metallic superconductivity arising from only electron-electron interactions. The basic physical idea is deceptively simple and appealing:  Plasmons are quantized bosonic collective modes of the well-known `plasma wave' charge density oscillations arising from the long-range Coulomb coupling, and the electron-plasmon interaction may act similarly to the electron-phonon interaction providing the pairing glue, leading to metallic superconductivity.  If so, the putative $T_c$ of such plasmon-induced superconductivity could be very high since the energy scales of all electronic excitations in metals are extremely high, $> 10^4$~K in contrast to phonons whose characteristic energy scale is $\sim 10^2$~K.  So, by just changing the bosonic glue from phonons to plasmons, it seems plausible that the $T_c$ could be raised by $> 2$ orders of magnitude!  In fact, the cuprate SCs with $T_c \sim 100$~K are thus heuristically consistent with an effective electron-electron induced SC since their $T_c$ is crudely a factor of 100 larger than the corresponding $T_c \sim 1$--$10$~K in conventional metals such as Al and Pb which are known to manifest phonon-mediated SC for sure.  This has motivated people to propose the plasmon mechanism for the cuprate superconductivity although this is definitely a rather small minority view by no means accepted in the larger high-$T_c$ community~\cite{Ruvalds1987,Ishii1993,Bill2003,Kresin1988}.  The crucial problem in such suggestions for plasmon mediated superconductivity in electron liquids is the stark absence of generic high-temperature ($T_c > 100$~K) metallic superconductivity in normal metals which all have plasmons with energy scales $> 10^4$~K and Fermi energies also $> 10^4$~K.  If plasmons and electron-electron interactions could generically lead to metallic superconductivity, Na and K and Li should all be room temperature superconductors, but factually, they are not superconductors at all under ambient conditions in spite of being ideal jellium electron liquids with electronic energy scales $\sim 5 \times 10^4$~K.  But, all known ($> 50$ different elements, some under high pressure) metallic superconductors with plasmon energies $> 10^4$~K, are known to be phonon-mediated $s$-wave SCs with $T_c$ (mostly) $\sim 1$--$10$~K which is in line with the typical phononic energy scale $\sim 10^2$~K.  In fact, a major problem with the plasmon-mediated (or other similar) electron interaction mediated SC proposal is:  Why is the $T_c$ so low in all such SCs (even in high-$T_c$ cuprates, $T_c \sim 100$~K, orders of magnitude below the electronic kinetic energy and interaction energy scales)?

We show in this work, using plausible BCS theory based arguments, that the generically expected BCS $T_c$ for the plasmon mediated metallic SC is $\sim 100$--$1000$~K, an absurdly high number when talking about conventional metallic superconductors which have been experimentally studied extensively for almost 120 years.  Obviously, this `plausible' straightforward BCS theory, with the plasmon modes replacing the phonon modes and the electron-plasmon coupling replacing the electron-phonon coupling, does not work as it predicts manifestly incorrect $T_c$ which is incompatible with reality.  Alkali metals are not $\sim 100$~K superconductors, and the superconducting metals get their superconductivity from electron-phonon interactions, not the electron-electron interactions.  Therefore, the question arises what is wrong with the BCS theory applied to plasmons compared with that for phonons where generally it has had predictive success in producing $T_c$ and superconducting gaps in agreement with experiments in numerous situations~\cite{Allen1983,Allen1999}.  We discuss this point from several critical perspectives using the Eliashberg theory, emphasizing that it is possible to predict essentially any $T_c$ based on the BCS-Eliashberg theory for plasmon-induced superconductivity since the theory is uncontrolled because of the crucial inapplicability of Migdal theorem for the electron-plasmon interaction~\cite{Migdal1958}.  The successful BCS-Eliashberg theory for superconductivity, enabling a reasonable quantitative $T_c$ prediction for metallic superconductors, is grounded on the Migdal theorem which applies to the electron-phonon interaction problem in metals by virtue of the Fermi energy being much larger than the typical phonon energy (the so-called Debye temperature), but not for the electron-electron interaction induced plasmonic superconductivity.

The Migdal theorem rules out all higher order vertex corrections for electron-phonon interaction induced superconductivity showing the vertex corrections to all orders going as $(m/M)^{1/2}$, where $m$ ($M$) are the electron (ion) mass, with $M/m > 2000$ in general.  Remarkably, this result for the smallness of vertex corrections is independent of the electron-phonon coupling strength, applying equally to weak-coupling, e.g., Al ($\lambda_{\text{ph}} < 1$) and strong-coupling, e.g., Pb ($\lambda_{\text{ph}} > 1$) metallic SCs.  The small parameter $m/M$ in the Migdal theorem asserting the negligible contributions by all vertex corrections is equivalent to the smallness of $v_{\text{ph}}/v_F$ or $\omega_D/E_F$ for metals, where $v_{\text{ph}}$ ($v_F$) are the sound velocity (Fermi velocity) and $\omega_D$ ($E_F$) are the Debye energy (Fermi energy).  The existence of Migdal theorem enables a rigorous quantitative theory, the BCS-Eliashberg-Migdal theory, for superconductivity, allowing predictions of $T_c$ based on microscopic parameters such as the electron-phonon coupling constant and the details of phonon and electron dispersion.  In addition, and this is another important aspect of key physics, the Eliashberg theory allows, in principle, the inclusion of Coulomb repulsion which reduces the effective electron-phonon coupling constant $\lambda_{\text{ph}}$ by a parameter universally referred to as $\mu^*$.  This parameter is typically estimated by approximately  solving the Eliashberg equation, and for most metals, $\mu^* \sim 0.15$, implying that there is no superconductivity if $\lambda_{\text{ph}} < 0.2$, which is for example the situation for Cu (or Na) which do not go superconducting as Coulomb repulsion negates the effective attractive coupling induced by phonons.

Now we immediately face a serious conundrum in discussing superconductivity induced by plasmons.  Since the plasmons arise intrinsically as collective modes of the interacting electron system, unlike the phonons which arise from the lattice vibrations, there cannot be, by definition, any Migdal theorem for the electron-plasmon interaction since there is no separation between electron and plasmon energy scales.  Indeed, the plasmon energy in metals ($\sim 10$~eV) is comparable to the Fermi energy $E_F$, and is also similar to the typical average metallic inter-electron Coulomb potential energy.  Thus, there is no energy scale separation (or velocity scale separation) enabling the neglect of vertex corrections. Hence, the use of BCS-Eliashberg theory for the calculation of $T_c$ in the plasmon induced metallic superconductivity is fundamentally flawed as one must include vertex corrections to all orders in the theory, which is obviously impossible. We show that it is easy to obtain $T_c \sim 100$--$1000$~K in metals arising from electron-plasmon interaction simply by following the standard BCS-Eliashberg-Migdal theory (with plasmons replacing phonons), but these results are unreliable since the neglect of vertex corrections in the theory is unjustified and uncontrolled.  Similarly, the estimate of $\mu^*$ in the plasmon case is nontrivial and ambiguous since one is starting entirely with a repulsive interaction.  We elaborate on these subtle but important theoretical issues in the rest of this paper.

We emphasize that the paper has two distinct parts, which are related conceptually, but disconnected quantitatively.  On the one hand, we show that any naive use of the BCS theory, even using reasonable Coulomb repulsion estimates, would invariably produce very high $T_c$ in jellium metals, implying that most metals including  Na, K etc. should be  high-$T_c$ superconductors due to the plasmon-mediated inter-electron attraction, which they most certainly are not.  The reason for this is mathematically simple:  The prefactor (in the pre-exponetial term) of the BCS formula has an energy scale determining the overall $T_c$ energy scale, which for phonon-mediated superconductivity  is typically the Debye energy ( $\sim 10^2\,K$), whereas for plasmon-mediated superconductivity this energy scale is the Fermi energy or the plasmon energy at Fermi momentum, which for 3D metals is $\sim 10^4\,K$.  Thus, simple considerations without vertex corrections invariably lead to very high $T_c$ in all theories on plasmon-mediated suoerconductivity, which is prima facie invalidated by experiments.  Regular metals are not plasmon-mediated superconductirs with $T_c\,100-1000K$, they are phonon mediated superconductors with $T_c\sim 1-10K$.  In fact, most alkali metals are not superconductors at all under pristine conditions.   This leads to the conclusion that the standard application of BCS-type theories to plasmon mediated superconductivity leads invariably to results which are empirically invalid  in real materials.  On the other hand, we also provide a detailed Eliashberg theory with quantitative results, but the theory is not definitive because of the cut off, which at this stage of the theory is somewhat arbitrary.  We show explicitly that the eventual $T_c$ depends on the cutoff.  In pronciple, this cutoff is associated with the suppression of $T_c$ by the Coulomb repulsion effect, which is difficult to ascertain quantitatively even for the phonon-induced superconductivity in metals.  The absence of a Migdal theorem in plasmon-induced superconductivity makes the situation far worse, but typically we believe that $T_c$ is low or even zero although this part of our work needs further improvement in the future including vertex corrections. As such, the two parts of our work are only indirectly connected.  Although we believe that repulsive Coulomb interaction does not produce any practical s-wave superconductivity in metals, the presence of the cut off dependence rules out our making any definitive statement about whether low $T_c$ superconductivity is allowed here.  But the first part of our work, establishing that any naive application of the BCS theory for plasmon-induced supercoductivity, produces absurdly high $T_c$ (and is therefore dubious) remains generally valid.  The strong-coupling situation, addressed in our dull Eliashberg theory remains technically challenging because of the appearance of the cutoff, separating the Migdal-valid low-frequency ($<$ cutoff) and the higher frequency regime where Migdal-Eloashberg theory no longer applies.  Since $T_c$ is invariably low (or vanishig) for geberally low cut off, our tentaive conclusion is that plasmon-mediated superconductivity generically may not arise in jellium metals, but this concusion is more an educated conjecture than a rigorous theoretical finding since we know of no way of solving the problem going beyond the Migdal-Eliashberg formalism.

We emphasize that we do not by any means claim to have solved the problem of plasmon-induced superconductivity in metals (or the more general electron-electron interaction induced superconductivity).  It is inherently a strong-coupling problem (even the KL SC is a leading-order perturbative result only) where the absence of vertex corrections makes all theoretical attempts ad hoc and questionable.  We do not know what the neg;ected higher order vertx terms would do as we know for the electron-phonon SC by virtue of the Migdal theorem. All we are doing in this work is to cast a serious question mark on all claims of an s-wave SC arising from plasmon-mediated interactions by pointing out several fundamental  theoretical shortcomings, which are typically not discussed in papers claiming the appearance of plasmon-mediated superconductivity.

The rest of this paper is organized as follows.  In Sec.~\ref{sec:coulomb}, we provide the calculated dynamically screened Coulomb interaction in 3D and 2D jellium metals for several many-body approximations to show explicitly that indeed the effective screened interaction is attractive in large regimes of the energy-momentum space, making a discussion of possible superconductivity arising strictly from electron-electron interactions meaningful.  In Sec.~\ref{sec:bcs}, we use the effective electron-plasmon interaction from Sec.~\ref{sec:coulomb} to obtain the SC $T_c$ induced by plasmons in the standard BCS theory, finding $T_c \sim 100$~K in typical metals (which is in stark disagreement with experiments).  In Sec.~\ref{sec:eliashberg-migdal}, we develop a full Eliashberg-Migdal theory, and point out precisely why the theory is unreliable in predicting $T_c$ by discussing the subtle frequency-dependent retardation effects which are crucial in the Eliashberg theory.  We conclude in Sec.~\ref{sec:conclusion}, summarizing our findings and emphasizing that no compelling theoretical case can be made that electron-plasmon interaction can or should lead to conventional and practical superconductivity in normal metals, although uncontrolled approximations do typically lead to $T_c \sim 100$~K for plasmon-induced metallic superconductivity.
Appendix \ref{sec:coulomb-2D} provides some of  the technical details on the screened interaction.

\section{Dynamically screened Coulomb interaction}\label{sec:coulomb}

\begin{figure*}
    \includegraphics[width=0.9\linewidth,trim={0 1cm 0 0},clip]{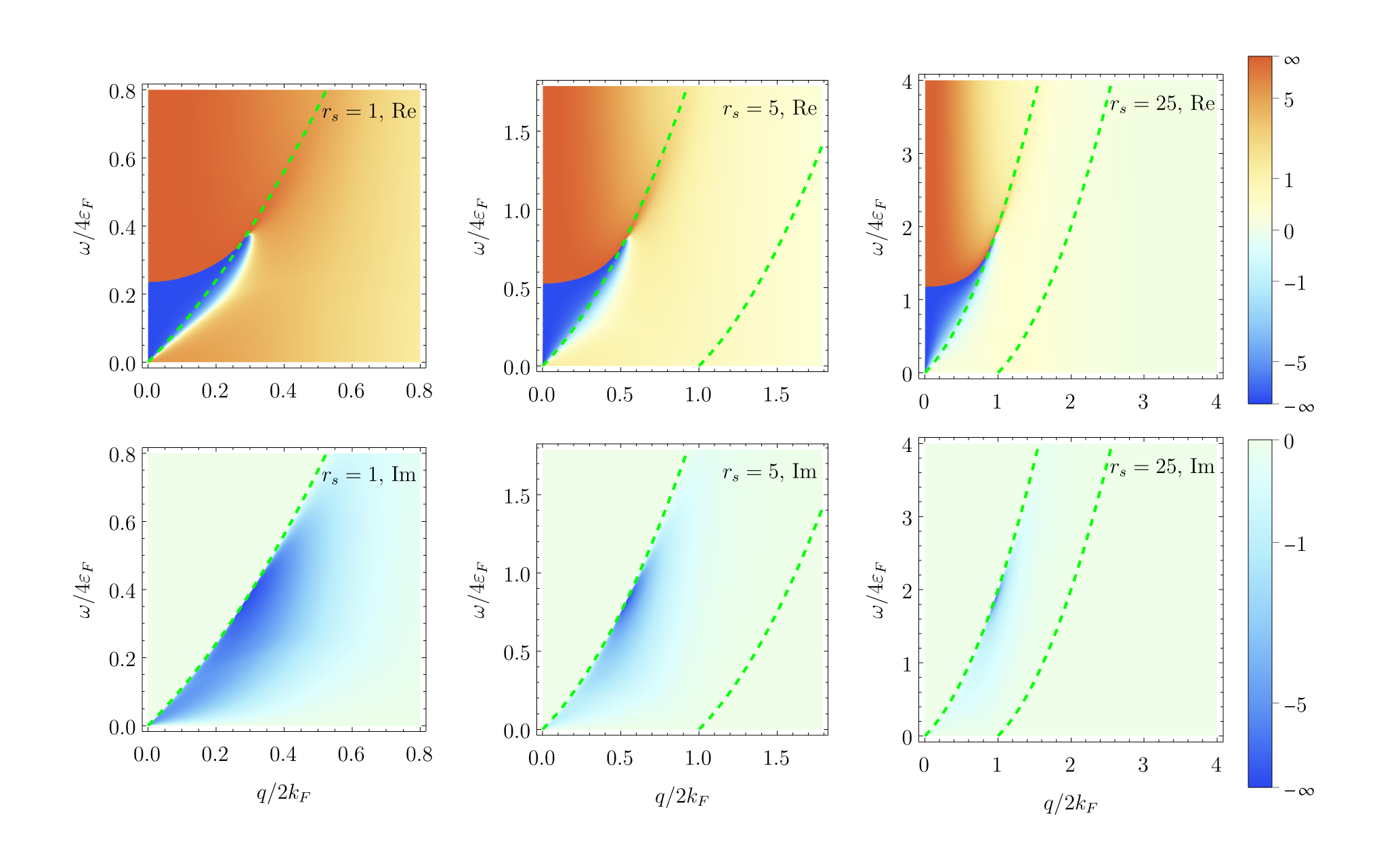}
    \caption{RPA Coulomb interaction $\tilde u_\text{RPA}(q,\omega)$ with various $r_s$ and the top (bottom) row showing the real (imaginary) part. Dashed lines indicate the damped region. Note the large region with negative real part and zero imaginary part, indicating an attractive interaction.}
    \label{fig:RPA}
\end{figure*}

In this section, we calculate and plot the dynamically screened Coulomb interaction in a three-dimensional electron gas (jellium metal) for several approximations to show that there is a large regime in the energy-momentum space where the effective Coulomb interaction is undamped and attractive. This shows that it is meaningful to discuss superconductivity appearing purely due to Coulomb interaction.

Consider a three-dimensional electron gas with dispersion $\varepsilon_\mathbf{p}=p^2/2m$ and spin degeneracy 2.
At zero temperature, the Lindhard polarizability~\cite{Lindhard1954}
\begin{equation}
    \chi(\mathbf{q},iq_n)=2\int \frac{d\mathbf{p}}{(2\pi)^3}\frac{f_\mathbf{p}-f_{\mathbf{p}+\mathbf{q}}}{iq_n+\varepsilon_\mathbf{p}-\varepsilon_{\mathbf{p}+\mathbf{q}}},
\end{equation}
can be calculated analytically ($f_\mathbf{p}=\theta(\varepsilon_\mathbf{p}-\mu)$).
Substitute $iq_n=\omega+i\epsilon,\epsilon\to 0$ for the retarded response and define dimensionless variables $\tilde q=q/2k_F$, $\tilde\omega=\omega/4\varepsilon_F$ and $\tilde\chi=\chi/d(\varepsilon_F)$ (where $d(\varepsilon_F)$ is the density of states at the Fermi energy $\varepsilon_F$, and $k_F$ is the corresponding Fermi momentum), we have~\cite{Bruus2004,Mahan1980}
\begin{equation}
    \operatorname{Re}\tilde\chi(\tilde q,\tilde\omega+i\epsilon)=-\frac{1}{2}-\frac{f(\tilde q,\tilde\omega)+f(\tilde q,-\tilde\omega)}{8\tilde q},
\end{equation}
\begin{equation}
    f(\tilde q,\tilde\omega)=\left[1-\left(\frac{\tilde\omega}{\tilde q}-\tilde q\right)^2\right]\,\ln\left|\frac{\tilde q+\tilde q^2-\tilde\omega}{\tilde q-\tilde q^2+\tilde\omega}\right|,
\end{equation}
\begin{multline}
    \operatorname{Im}\tilde\chi(\tilde q,\tilde\omega+i\epsilon)=\\
    -\begin{cases}
        \frac{\pi}{8\tilde q}\left[1-\left(\frac{\tilde\omega}{\tilde q}-\tilde q\right)^2\right] & \text{if }|\tilde q-\tilde q^2|<\tilde\omega<\tilde q+\tilde q^2, \\
        \frac{\pi}{2}\frac{\tilde\omega}{\tilde q} & \text{if }0<\tilde\omega<\tilde q-\tilde q^2, \\
        0 & \text{otherwise}.
    \end{cases}
\end{multline}

Under the random phase approximation (RPA), the dielectric function is approximated by~\cite{Mahan1980,Hwang2018,schrieffer2018theory,Fetter2003}
\begin{equation}\label{eq:epsilon-RPA}
    \varepsilon_\text{RPA}(\mathbf{q},\omega)=1-v_c(q)\chi(\mathbf{q},\omega+i\epsilon)
\end{equation}
where $v_c(q)=4\pi e^2/q^2$ is the bare 3D Coulomb coupling.
The screened Coulomb interaction is
\begin{equation}
    u_\text{RPA}(\mathbf{q},\omega)=\frac{v_c(q)}{\varepsilon_\text{RPA}(\mathbf{q},\omega)}
\end{equation}
Defining the dimensionless interaction $u=\frac{\pi e^2}{k_F^2}\tilde u$, we can express everything using dimensionless parameters as
\begin{equation}
    \tilde u_\text{RPA}(\tilde q,\tilde\omega)=\frac{1}{\tilde q^2-\frac{1}{\pi}\left(\frac{9\pi}{4}\right)^{-1/3} r_s \tilde\chi(\tilde q,\tilde\omega+i\epsilon)},
\end{equation}
where $r_s$ is the well-known dimensionless Wigner-Seitz radius, which is simply a measure of the relative strength of the Coulomb coupling compared with the zero point energy of the electrons (i.e.\ the Fermi energy), defined universally as the average distance between the electrons in the units of the effective Bohr radius: $r_s= (me^2/\hbar^2) (3/4\pi n)^{1/3}$

The function $\tilde u_\text{RPA}(\tilde q,\tilde\omega)$ is plotted in Fig.~\ref{fig:RPA} for various values of $r_s$, with the region bounded by the two dashed lines, corresponding to $\tilde q^2-\tilde q<\tilde\omega<\tilde q+\tilde q^2$, being the electron-hole continuum where the screened Coulomb interaction is damped (having a negative imaginary part).
From the figure, it is clear that there is a large, attractive interaction regime indicated by $\operatorname{Re}u(q,\omega)<0$, $\operatorname{Im}u(q,\omega)=0$ (the blue color in the real part plots), between the left boundary of the damped regime (the left of the first dashed line) and the plasmon pole (the boundary between blue and orange).
The fact that the dynamically screened interaction is actually attractive in large regimes of the $(q,\omega)$ space is unexpected and highly counter-intuitive, but it is nevertheless true.

\begin{figure*}
    \includegraphics[width=0.9\linewidth,trim={0 1cm 0 0},clip]{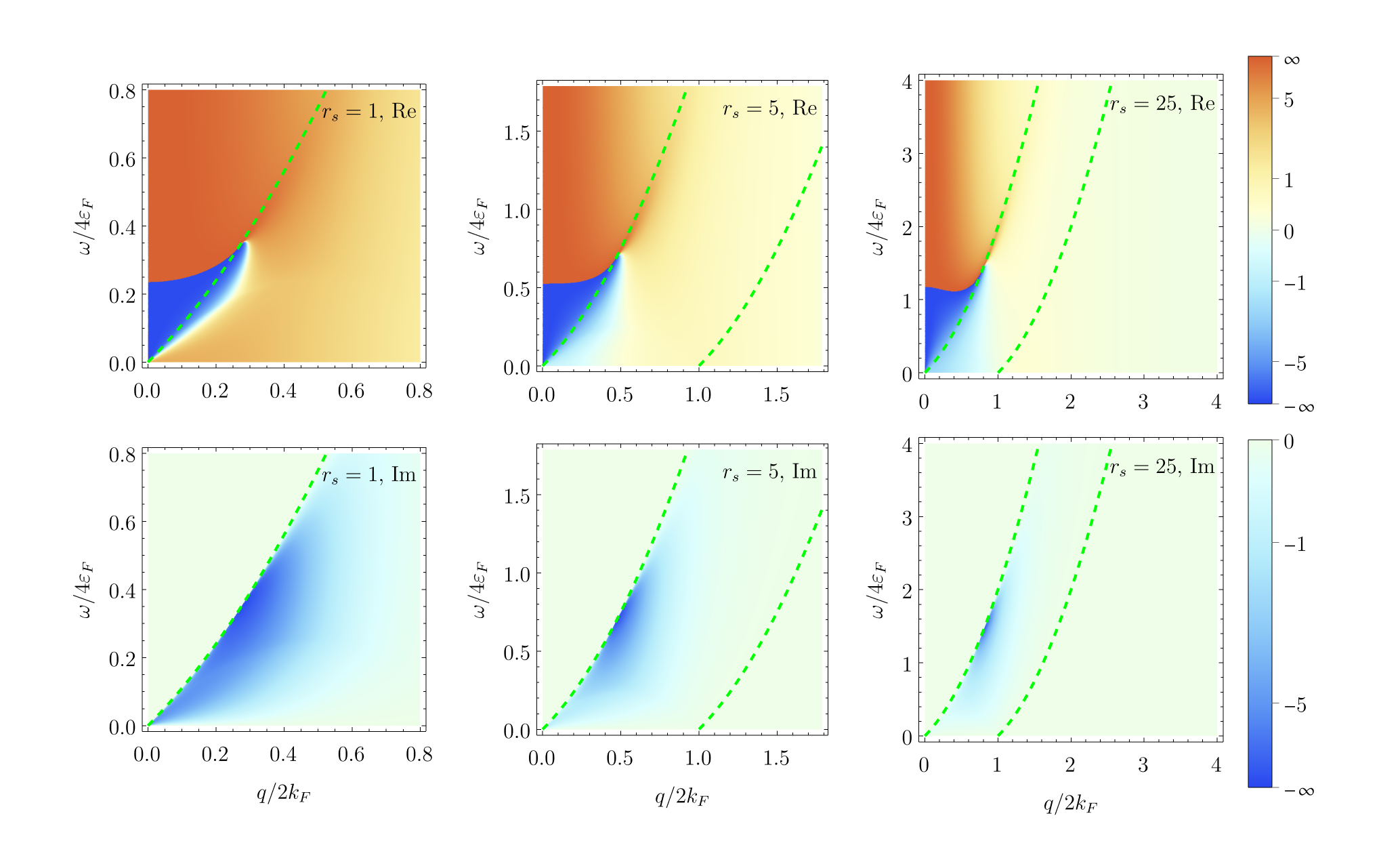}
    \caption{Hubbard-corrected RPA Coulomb interaction $\tilde u_\text{H}(q,\omega)$ with various $r_s$ and the top (bottom) row showing the real (imaginary) part. Dashed lines indicate the damped region, which is identical to that of RPA. Note the qualitative similarity to Fig.~\ref{fig:RPA}.}
    \label{fig:Hubbard}
\end{figure*}

One may think that this might be an artifact of RPA, although RPA is the only available controlled approximation for the dielectric function, being a systematic expansion in $r_s$ summing up the most divergent ring diagrams in each order~\cite{Fetter2003}.
Here, we use the Hubbard local field correction to the RPA~\cite{Hubbard1957}. That is, instead of Eq.~(\ref{eq:epsilon-RPA}), the dielectric function is now approximated by~\cite{Mahan1980}
\begin{equation}\label{eq:epsilon-Hubbard}
    \varepsilon_\text{H}(\mathbf{q},\omega)=1-\frac{v_c(q)\chi(\mathbf{q},\omega+i\epsilon)}{1+v_c(q)G(q)\chi(\mathbf{q},\omega+i\epsilon)},
\end{equation}
where the correction factor is
\begin{equation}
    G(q)=\frac{1}{2}\frac{q^2}{q^2+k_F^2}=\frac{1}{2}\frac{1}{1+(\frac{1}{2\tilde q})^2}.
\end{equation}
The resulting dimensionless Coulomb interaction $\tilde u_\text{H}(\tilde q,\tilde\omega)$ is plotted in Fig.~\ref{fig:Hubbard}.
One can see that although the shape of the attractive regime is distorted a bit, especially at large $r_s$, no qualitative result is changed.
(Note that as $G(q)$ is real, it does not modify the shape of the damped regime.)
This is expected as the correction factor $G(q)$ is only effective when $q$ is at least comparable to $k_F$, the attractive regime, which is mostly below $k_F$, only gets slightly modified.
There are many proposed local field corrections in which $G(q)$ has a more complicated form. However, we do not expect the results to change much, as such corrections are typically not effective when $q$ is much smaller than $k_F$ (where much of the large attraction regime is in).
In any case, all these local field corrections are invariably static, and are thus unable to shed any light on the crucial retardation effects arising from the frequency dependence of the effective interaction relevant for superconductivity.  In addition, the theory of local field corrections is uncontrolled from a diagrammatic perturbation viewpoint, often mixing orders (and thus violating Ward identities), and is thus less reliable than RPA from a conceptual viewpoint.  There are occasional claims in the literature that such local field corrections are equivalent to including vertex corrections in the theory which RPA neglects.  This is indeed true. The local field corrections through a phenomenological $G(q)$ incorporated in the RPA dielectric function, as in the Hubbard approximation above, are indeed including vertex corrections in a crude frequency-independent ad hoc manner, but in all likelihood such arbitrary vertex corrections make the theory worse because it neglects frequency dependence and certainly violates Ward identities since various diagrammatic orders are arbitrarily mixed in such ad hoc approximations.  The important point to emphasize in the context of the current work is that including a $G(q)$ or not in the dielectric function makes no significant change in our results and conclusions whatsoever.

\begin{figure*}
    \includegraphics[width=0.9\linewidth,trim={0 1cm 0 0},clip]{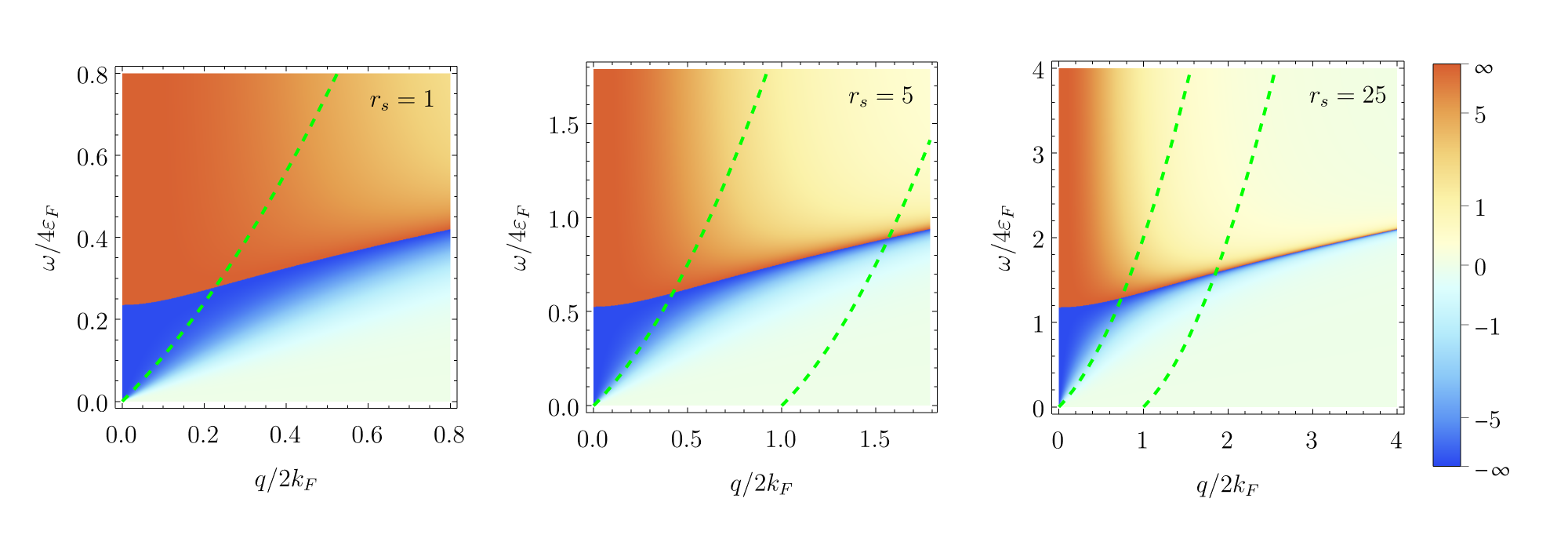}
    \caption{Plasmon-pole approximated Coulomb interaction $\tilde u_\text{PPA}(q,\omega)$, with dashed line indicate the damped region from RPA. Note the similarity to Fig.~\ref{fig:RPA} in the region to the left of the first dashed line, indicating that the large attractive region is essentially due to the plasmon pole.}
    \label{fig:PPA}
\end{figure*}
\begin{figure*}
    \includegraphics[width=0.9\linewidth,trim={0 1cm 0 0},clip]{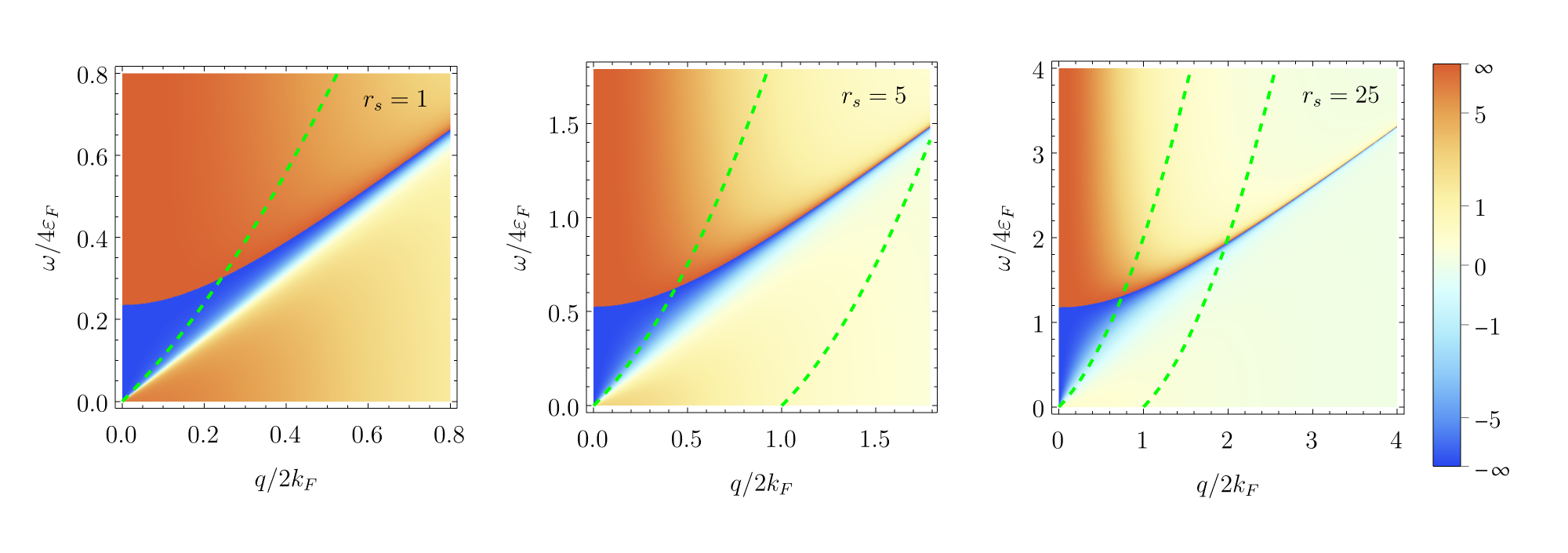}
    \caption{Hydrodyanmic approximated Coulomb interaction $\tilde u_\text{hydro}(q,\omega)$, with dashed line indicate the damped region from RPA. As in \ref{fig:PPA}, it reproduces the essential feature of Fig.~\ref{fig:RPA} in the region to the left of the first dashed line.}
    \label{fig:hydro}
\end{figure*}

Since the large attractive regime in the energy-momentum space (the blue color in the real part plots) is directly extended from the plasmon pole (the line between blue and orange), a natural question is whether the plasmon pole approximation (PPA) itself is enough to reproduce the same qualitative result.
The PPA is an approximation to the RPA where the response is represented entirely by a pole in the real part of response function (so the imaginary part becomes a delta function at this plasmon pole).  The PPA is used extensively in the literature and is known to reproduce RPA results quantitatively~\cite{Lundqvist1967,Lundqvist1967a,Overhauser1971}.
Under the plasmon pole approximation, the dielectric function is~\cite{Bruus2004}
\begin{equation}
    \varepsilon_\text{PPA}(\mathbf{q},\omega)=1-\frac{\omega_p^2}{\omega^2}\left[1+\frac{3}{5}\left(\frac{qv_F}{\omega}\right)^2\right],
\end{equation}
where $\omega_p$ is the 3D long wavelength plasmon frequency
\begin{equation}\label{eq:omega-p}
    \omega_p=\sqrt{\frac{4\pi n e^2}{m}}=(4\varepsilon_F)\cdot0.2351\sqrt{r_s},
\end{equation}
$n$ is the 3D electron density, $v_F$ is the Fermi velocity, and we have used the unit of $\hbar=1$. The resulting dimensionless Coulomb interaction $\tilde u_\text{PPA}(\tilde q,\tilde\omega)$ is plotted in Fig.~\ref{fig:PPA}.
One can see that it indeed reproduces the same attractive regime as the RPA case in Fig.~\ref{fig:RPA}, except that we must cut it off by the damped regime by hand, since PPA does not include the imaginary part (with the imaginary part simply being a delta function at the plasmon pole).
Alternatively, one can use the hydrodynamic approximation~\cite{DasSarma1979,DasSarma1982a,Hwang2018}, where the dielectric function is
\begin{equation}
    \varepsilon_\text{hydro}(\mathbf{q},\omega)=1-\frac{\omega_p^2}{\omega^2-\frac{3}{5}v_F^2q^2}.
\end{equation}
The resulting dimensionless Coulomb interaction $\tilde u_\text{hydro}(\tilde q,\tilde\omega)$ is plotted in Fig.~\ref{fig:hydro}.
Again, it reproduces the essential results contained in Fig.~\ref{fig:RPA}.

The results above clearly show that there is indeed a large regime in the energy-momentum space where the dynamically screened repulsive Coulomb interaction of the electrons produces an effective attractive interaction, making the discussion on that whether it leads to a pairing instability relevant.

\begin{figure*}
    \includegraphics[width=0.9\linewidth,trim={0 1cm 0 0},clip]{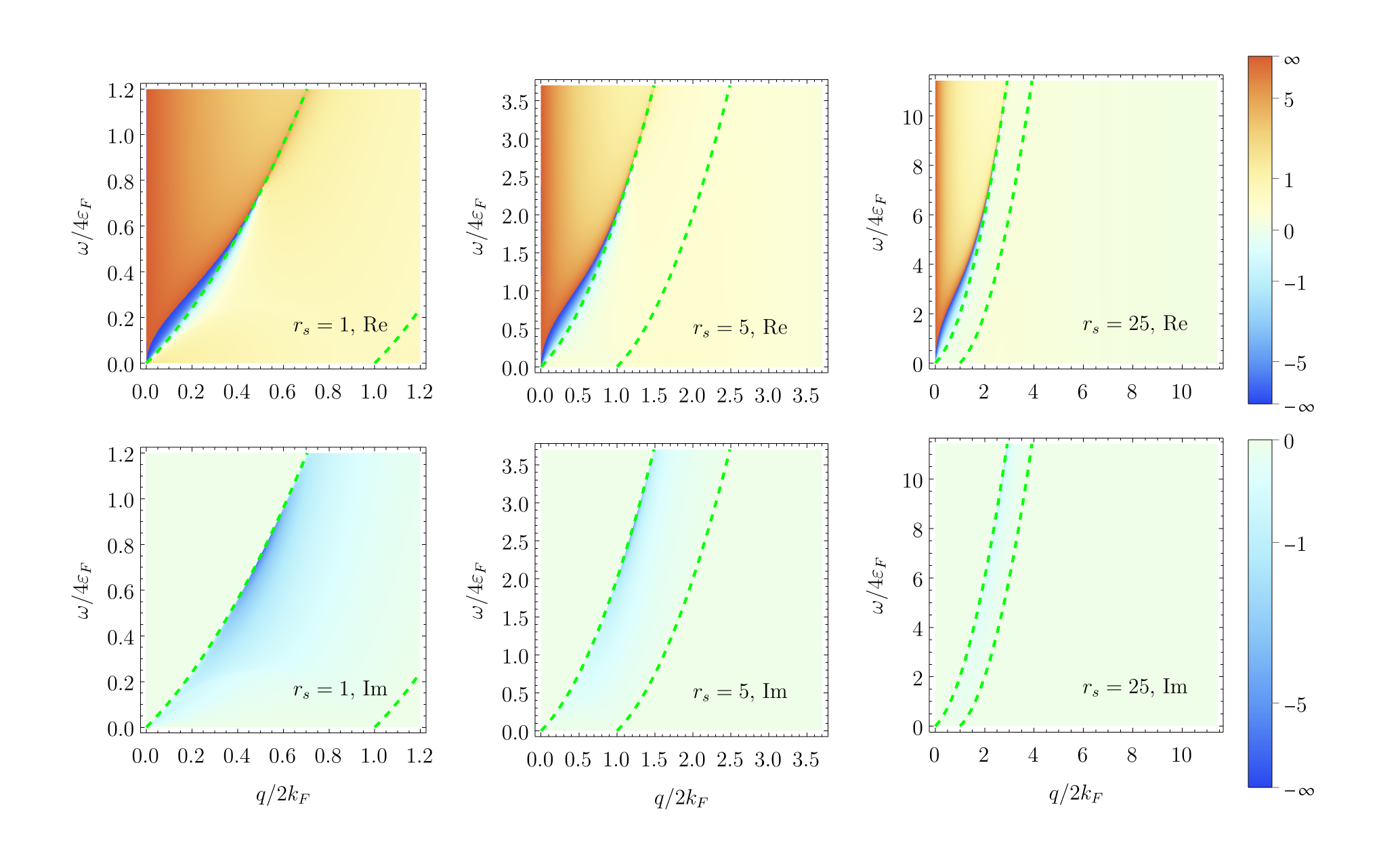}
    \caption{2D RPA Coulomb interaction $\tilde u^\text{2D}_\text{RPA}(q,\omega)$ with various $r_s$ and the top (bottom) row showing the real (imaginary) part. Dashed lines indicate the damped region. Note the different behavior of the plasmon dispersion from 3D.}
    \label{fig:2D_RPA}
\end{figure*}

\begin{figure*}
    \includegraphics[width=0.9\linewidth,trim={0 1cm 0 0},clip]{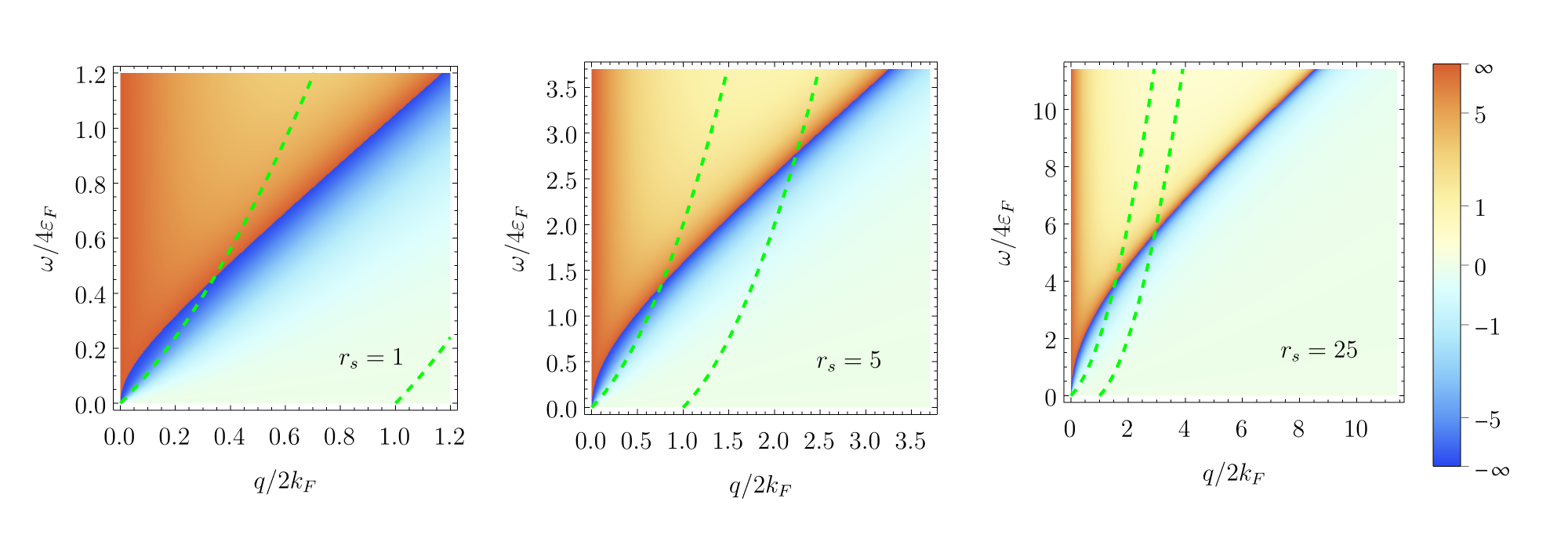}
    \caption{2D plasmon pole approximated Coulomb interaction $\tilde u^\text{2D}_\text{PPA}(q,\omega)$, with dashed line indicate the damped region from RPA. It reproduces the essential feature of Fig.~\ref{fig:2D_RPA}}
    \label{fig:2D_PPA}
\end{figure*}

We have also carried out similar calculations for the 2D electron gas in a jellium model (only for RPA and PPA), and the numerical results are shown in Figs.5 and 6 (for 2D RPA and PPA, respectively).  We provide the 2D analytical results in the Appendix~\ref{sec:coulomb-2D}.  Again, RPA and PPA give similar results in the 2D case, demonstrating that the plasmon pole approximation captures the basic physics of dynamical screening very well.  There is one glaring quantitative difference between 3D (Figs.~\ref{fig:RPA} and \ref{fig:PPA}) and 2D (Figs.~\ref{fig:2D_RPA} and \ref{fig:2D_PPA}).  The attractive regions (in blue in all figures) in the momentum (wavevector)-energy (frequency) space for the screened interaction is drastically smaller in the 2D case than in the 3D case---the interaction is notably less attractive in 2D compared with 3D.  This difference arises from the fact that the long wavelength, $q=0$,  plasma frequency vanishes in 2D as $q^{1/2}$ since the 2D plasmon dispersion goes as $\omega_p(q) = (2\pi n e^2 q/m)^{1/2}$ in the leading order (with $n$ as the 2D metallic electron density)~\cite{Stern1967}, in contrast to the constant $q$-independent 3D plasma frequency.  This considerably shrinks the regime between the plasmon branch $\omega_p(q) \sim q^{1/2}$ and the electron-hole continuum $\sim q$ before the plasmon hits the electron-hole continuum becoming damped.  Since the screened interaction can only be negative or attractive for $\omega<\omega_p$, naturally the attractive region shrinks in 2D since the plasmon energy is bounded by the $q^{1/2}$ dispersion.  Other than this quantitative difference, qualitatively 2D and 3D plots for the dynamically screened interactions are similar.
One naive implication of this difference between 2d and 3D dynamically screened Coulomb interaction may, however, be relevant to the consideration of Coulomb interaction induced SC.  This nominally implies that plasmon-induced SC is far less likely in 2D than in 3D since the effective electron-electron interaction is much less attractive (or more precisely, the phase space over which the interaction is attractive is much reduced) in 2D compared with 3D.  In fact, a similar, but somewhat subtle, situation arises in the KL SC phenomena also, where naive arguments lead to a nonexistence of KL SC in 2D~\cite{Chubukov1993}.

In the next two sections, we discuss whether the attractive effective interaction shown in Figs.~\ref{fig:RPA} and \ref{fig:Hubbard} could lead to conventional $s$-wave superconductivity with observable $T_c$ in normal metals using the simple BCS theory (Sec.~\ref{sec:bcs}) and the full Migdal-Eliashberg theory (Sec.~\ref{sec:eliashberg-migdal}).

\section{BCS theory for plasmon-induced metallic superconductivity}\label{sec:bcs}

The BCS theory is the paradigmatic theory for superconductivity with huge quantitative success for phonon-induced superconductivity in thousands of materials including all elemental metals (sometimes in the form of its strong-coupling extensions).  Since the BCS theory is textbook material, we do not provide any redundant details, see, e.g., Refs.~\cite{schrieffer2018theory,Fetter2003,Mahan1980} (Details are provided in the next section, Sec.~\ref{sec:eliashberg-migdal}, where the rigorous Eliashberg-Migdal theory is discussed in the context of electron-electron interaction induced superconductivity.)  In this section, we assume the applicability of the textbook BCS theory, and discuss its implications for $T_c$ in the plasmon induced superconductivity scenario of  3D jellium electron liquids.
Most, if not all, existing theories of the plasmon-induced SC in metals basically follow the simple BCS prescription. in some form or other.

The BCS theory provides the following formula for $T_c$ for boson mediated SC, where the typical boson energy is $\omega_b$ and $\lambda_b$ is the dimensionless electron-boson coupling producing the SC:
\begin{equation}\label{eq:BCS-Tc}
    T_c\sim\omega_b e^{-\frac{1}{\lambda_b}},
\end{equation}
where we have neglected an unimportant constant of $O(1)$ in the prefactor. (We use units $\hbar=k_B=1$ throughout unless otherwise explicitly noted, so temperature/frequency/energy and momentum/wavenumber/inverse wavelength are the same in our notation.)  For acoustic phonons, which are the primary boson-mediated mechanism driving SC in metals, the typical bosonic frequency $\omega_b\sim \omega_D$, where $\omega_D$ is the characteristic Debye energy for the phonons involved in the SC.   The dimensionless electron-phonon coupling $\lambda_b$ in Eq.~(\ref{eq:BCS-Tc}) varies between $0.1$ (Cu) to $1.7$ (Pb), and is defined by $\lambda_b= N(E_F)V$, where $N$ and $V$ respectively denote the electronic density of states at the Fermi level and $V$ the dimensionful electron-phonon coupling strength.  All we need to do is to find the corresponding expressions for the parameters $\omega_b$ and $\lambda_b$ for the plasmons in the electron-plasmon superconductivity problem with the plasmons being the bosonic glue producing pairing.

The simplest way to see a direct formal connection between the electron-phonon and the electron-plasmon interaction problems is to write down the dynamical (or the correlation) part of the  $T=0$  electron self-energy $M_c$ arising from the electron-electron interaction in the plasmon-pole approximation where the electronic response is characterized by a single plasmon pole instead of the full RPA dielectric function:
\begin{multline}\label{eq:BCS-Mc}
    M_c(k,\omega)=\int\frac{d^3 q}{(2\pi)^3}\frac{4\pi e^2}{q^2}\frac{\omega^2}{2\omega_q}\\
    \cdot\left(\frac{\theta(k_F-|\mathbf{k}-\mathbf{q}|)}{\omega-\varepsilon_{\mathbf{k}-\mathbf{q}}+\omega_q}+\frac{\theta(|\mathbf{k}-\mathbf{q}|-k_F)}{\omega-\varepsilon_{\mathbf{k}-\mathbf{q}}-\omega_q}\right).
\end{multline}
Here $\omega_p$ is the 3D long wavelength plasmon frequency given by Eq.~(\ref{eq:omega-p}). The integral is over the 3D wavevector space constrained by the 3D Fermi sphere with the Fermi wavevector of $k_F$.
The quantities $\varepsilon_k$ and $\omega_q$ denote the effective noninteracting single-particle electron dispersion and the effective plasmon pole energy, typically determined by the f-sum rule and static screening.  It is well-known that the plasmon-pole approximation represents an excellent quantitative approximation to the full RPA theory in 3D metals in spite of the apparent drastic approximation of neglecting the branch cut in the full RPA dynamical screening.  Note that there is an additional static (i.e.\ frequency-independent) exchange correction $M_x$ to the self-energy which is given by:
\begin{equation}\label{eq:BCS-Mx}
    M_x(k)=\int\frac{d^3 q}{(2\pi)^3}\frac{4\pi e^2}{q^2}\theta(k_F-|\mathbf{k}-\mathbf{q}|).
\end{equation}
The advantage of expressing the frequency dependent self-energy in the form of Eq.~(\ref{eq:BCS-Mc}) becomes heuristically obvious when we write the corresponding electron-phonon self-energy for 3D electrons interacting with phonons:
\begin{multline}\label{eq:BCS-Mph}
    M_\text{ph}(k,\omega)=\int\frac{d^3 q}{(2\pi)^3}\frac{|\gamma|^2}{2\omega_{\text{ph},q}}\\
    \cdot\left(\frac{\theta(|\mathbf{k}-\mathbf{q}|-k_F)}{\omega-\varepsilon_{\mathbf{k}-\mathbf{q}}-\omega_{\text{ph},q}}+\frac{\theta(k_F-|\mathbf{k}-\mathbf{q}|)}{\omega-\varepsilon_{\mathbf{k}-\mathbf{q}}+\omega_{\text{ph},q}}\right).
\end{multline}
We note that the plasmon-pole frequency $\omega_q$ here us essentially the finite q dispersion of the plasma frequency $\omega_p$ given by $\omega_q = \omega_p [ 1+ O(q^2 ) or O (q)]$ depending on 3D or 2D respectively. [16]   (Changing $\omega_q$ to just $\omega_p$ hardly changes anything in the theory.)
Here, $\omega_{\text{ph},q}$ and $\gamma$ are the phonon frequency and the electron-phonon coupling, respectively.  We emphasize the obvious apparent fact that Eqs.~(\ref{eq:BCS-Mc}) and (\ref{eq:BCS-Mph}) are formally identical with $\omega_{\text{ph},q}$ in Eq.~(\ref{eq:BCS-Mph}) being replaced by $\omega_q$ in Eq.~(\ref{eq:BCS-Mc}) and the electron-phonon coupling $\gamma^2$ in Eq.~(\ref{eq:BCS-Mph}) being replaced by the effective electron-plasmon coupling, $\gamma_p$, defined by:
\begin{equation}
    \gamma_p^2=v_c(q)\omega_p^2,
\end{equation}
where $v_c(q)=4\pi e^2/q^2$ is the 3D Coulomb coupling.

This precise formal mapping between the electron-phonon and electron-plasmon self-energies enables an immediate heuristic solution for the appropriate $T_c$ in the BCS theory for the electron-plasmon interaction, given by Eq.~(\ref{eq:BCS-Tc}), where the $\omega_b$ and $\lambda_b$ should now represent $\omega_p$ and $\lambda_p$, respectively, for the corresponding electron-plasmon interaction as described above.  This heuristic mapping between phonons and plasmons enables to express the plasmon-induced $T_c$ to be:
\begin{equation}\label{eq:BCS-Tcp}
    T_{cp}\sim\omega_p e^{-\frac{1}{\lambda_p}},
\end{equation}
with $\lambda_p$ given by:
\begin{equation}\label{eq:BSC-lambda-p}
    \lambda_p=\frac{e^2q_p}{\pi\omega_p},
\end{equation}
where $q_p=\omega_p/v_F$, with $v_F$ denoting Fermi velocity, is the cutoff momentum where the plasmon dispersion enters the electron-hole continuum  (see Sec.~\ref{sec:coulomb} and Fig.~\ref{fig:RPA}), becoming damped.
The plasmon no longer exists for $q>q_p$.  Note that Eq.~(\ref{eq:BSC-lambda-p}) is consistent with the estimates for the electron-plasmon coupling constant in other contexts~\cite{Antropov1993}.

\begin{figure}
    \centering
    \includegraphics[width=0.75\linewidth]{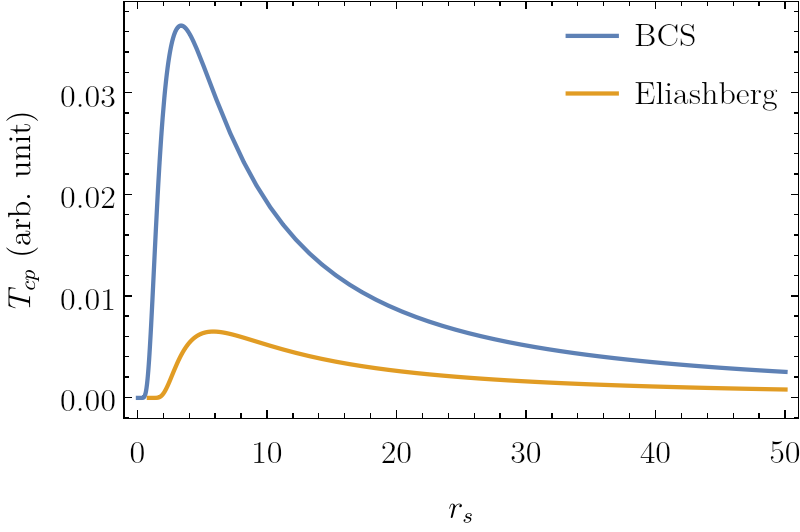}
    \caption{The scaling of the $T_c$ of plasmon-induced superconductivity with $r_s$.}
    \label{fig:Tcp}
\end{figure}

We now convert our results in terms of the dimensionless Wigner-Seitz radius $r_s$ (see Sec.~\ref{sec:coulomb}).
This provides the following scaling law for $T_{cp}$
\begin{equation}\label{eq:Tcp-scaling}
    T_{cp}\sim r_s^{-\frac{3}{2}} e^{-\frac{5}{r_s}}.
\end{equation}
In Eq.~(\ref{eq:Tcp-scaling}), the $r_s^{-3/2}$ comes simply from the dependence of the plasmon energy on density: $\omega_p \sim n^{1/2}$ with $n \sim 1/r_s^3$ (in 3D) by definition. Eq.~(\ref{eq:Tcp-scaling}) with proper prefactors, which would be a model-dependent number, is the basic BCS prediction for plasmon induced $T_c$ in metals.  The resultant $T_c$ (Fig.~\ref{fig:Tcp}) has a peak around $r_s\sim 1$, and it vanishes quickly at high density (small $r_s$ because the basic coupling strength decreases as $r_s$ decreases producing an exponential drop in $T_c$) and slowly at low density (large $r_s$ because the plasmon frequency prefactor vanishes in a power law for large $r_s$ or low density as $n^{-1/2}$).

We have also carried out the BCS $T_c$ calculation in 2D following the same procedure as for the 3D case described above.  The main differences between 2D and 3D are that in 2D the plasmon dispersion vanishes in the $q=0$ long wavelength limit and the 2D density of states and the static Thomas-Fermi screening wavevector $q_\text{TF}$ are constant independent of density:
$\omega_p(q) = (2\pi n e^2 q/m)^{1/2}$  in 2D in contrast to $\omega_p= (4\pi n e^2/m)^{1/2}$ in 3D, and $q_\text{TF} =2m e^2/\hbar^2$ in 2D in contrast to 
$q_\text{TF}=(4m e^2 (3\pi^2 n)^{1/3}/(\pi^2 \hbar^2))^{1/2}$ in 3D, with $n$ being the 2D or 3D electron density as appropriate.  These differences, however, do not affect the actual $T_c$ formula in the 2D case which is still given by the same equation as in the 3D case (i.e.\ Eq.~(\ref{eq:Tcp-scaling}) above) except that the dimensionless coupling parameter $r_s$ is now defined for the 2D case as 
$r_s =(\pi a_0^2 n)^{-1/2}$, where $a_0=\hbar/e^2m$ is the Bohr radius and $n$ is the 2D electron density.  Most of our discussions in this article  are for 3D systems, but exactly the same considerations apply in the 2D case too.

\begin{figure}
    \centering
    \includegraphics[width=0.75\linewidth]{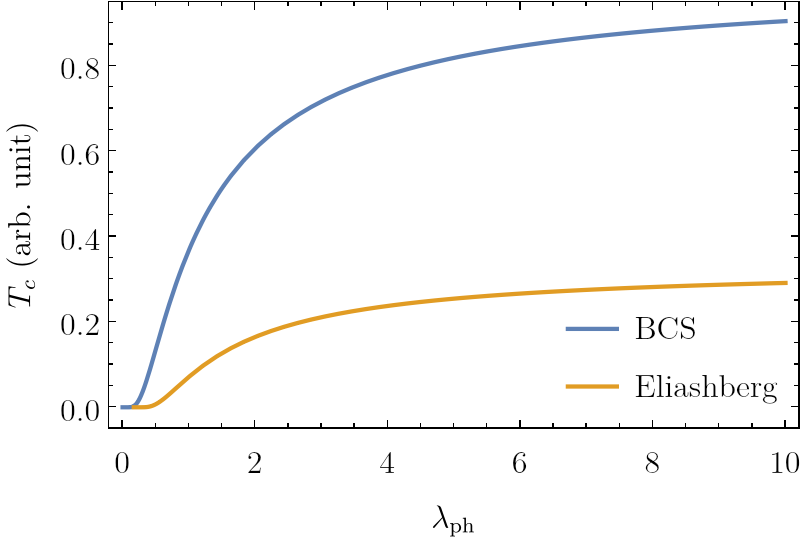}
    \caption{The scaling of the $T_c$ of phonon-induced superconductivity with $\lambda_\text{ph}$ (unit not comparable with Fig.~\ref{fig:Tcp}).}
    \label{fig:Tc}
\end{figure}

For the sake of comparison with the corresponding phonon-induced SC in metals (2D or 3D), we provide in Fig.~\ref{fig:Tc} the plots for the well-known BCS-Eliashberg formula for $T_c$ in metals as a function of the dimensionless electron-phonon coupling constant $\lambda_\text{ph}$ in units of the phonon frequency $\omega_\text{ph}$ assuming the accepted $\mu^*=0.15$ for the Eliashberg case~\cite{Allen1975}:
\begin{align}
    \text{BCS: }T_c &\sim \omega_\text{ph}e^{-\frac{1}{\lambda_\text{ph}}}\\
    \text{Eliashberg: }T_c &\sim \omega_\text{ph}e^{ -\frac{ 1 + \lambda_\text{ph}}{ \lambda_\text{ph} - \mu^* ( 1 +0.62  \lambda_\text{ph})}}\label{eq:Eliashberg-Tc}
\end{align}
Note that for $\mu^*=0$ and $\lambda_\text{ph}\ll 1$, the BCS and Eliashberg formula coincide as they must since the BCS theory is a weak-coupling ($\lambda_\text{ph}\ll 1$)  theory neglecting Coulomb repulsion ($\mu^*=0$). The main message from Figs.~\ref{fig:Tcp} and \ref{fig:Tc} are that the phonon-induced SC increases with the coupling eventually saturating at the effective phonon frequency for large coupling, but the plasmon-induced SC behaves qualitatively differently showing a maxima for an intermediate $r_s$ decreasing (exponentially) for $r_s\ll 1$ and in a power law manner for $r_s\gg 1$.  This nonmonotonicity in the plasmon SC arises from the fact that the plasmons are inherent properties of the electrons themselves and not an independent bosonic mode as in the phonon case.

We note that the current section deals with the weak-coupling BCS theory where the coupling glue (ie plasmons here) is weak, $\lambda<1$, so one should think of Eq. (22) above as a heuristic borrowed from the strong coupling theory~\cite{Allen1975}.  The point we are making here is that while the weak-coupling  BCS formula manifests a large peak in $T_c$ (see Fig. 7), the corresponding strong-coupling theory manifests a very weak peak.  We also note that there are some differences between the phonon (Fig. 8) and the plasmon (Fig. 7) case.  In the phonon case, $T_c$ keeps on growing slowly with increasing coupling, but in the plasmon case it effectively saturates after a weak peak in the strong coupling regime if Eq. (22) for the strong-coupling formula is used.  How real these differences are cannot be definitively decided ithout a true strong coupling theory for the plasmon case in the presence of vertex corrections.

What about the magnitude of the plasmon induced metallic superconductivity?  This is more easily seen by rewriting the transition temperature as $T_{cp} \sim \omega_p  \exp(-5/r_s)$, which makes it explicit that the scale of $T_{cp}$ is set by the plasmon energy (just as it is set by the Debye energy for the phonon induced superconductivity).  For normal metals, $r_s \sim 2$--$7$ whereas $\omega_p \sim 10^4$~K, leading to a typical metallic $T_{cp} \sim 1000$~K!  This is of course an absurd number, and explicitly demonstrates the problem with theories on plasmon induced superconductivity.  (By contrast, exactly the same BCS formula for phonon-induced metallic superconductivity, with $\lambda_b \sim 0.1$--$2$ and the typical Debye energy being $\sim 10^2$--$10^3$ gives the typical phonon induced metallic $T_c$ to be a reasonable $\sim 10$~K.)  Obviously, the naive use of the BCS theory to the plasmon-induced superconductivity leads to absurd room temperature superconductivity across the board for all normal metals, just by virtue of the plasmon (i.e.\ the electron-electron interaction) energy scale being too high, and the coupling strength (i.e.\ $r_s$) not being too small.

What are the plausible ways out of this conundrum of an absurd theoretical prediction of $T_{cp}>100$~K in normal metals, which is patently invalid?  We emphasize that simple fixes like improving the prefactor or changing the plasmon-pole approximation (or including local field corrections) only make $O(1)$ corrections, and cannot suppress $T_{cp}$ substantially, so the resolution is not in a somewhat improved BCS theory along the same lines.  For instance, using the full RPA interaction propagator replacing the plasmon-pole approximation does virtually nothing and the same is true for including local field corrections in the polarizability function.  In fact, such approximations are equally likely to enhance $T_{cp}$ slightly as they are to suppress it slightly.  Clearly, something radical is necessary, given the absurdity of the predicted $T_{cp}\sim 1000$~K. 

One possibility is the well-known Coulomb repulsion effect (the ``$\mu^*$ effect'')~\cite{Allen1975}, which has been already used in our Eq.~(\ref{eq:Eliashberg-Tc}) above for the phonon-mediated SC suppressed by Coulomb repulsion---this formula in Eq.~(\ref{eq:Eliashberg-Tc}) is the well-accepted Allen-Dynes empirical formula for metallic SC.  The direct Coulomb repulsion between the electrons, e.g.\ the frequency independent exchange self-energy of Eq.~(\ref{eq:BCS-Mx}) which depends only on the direct Coulomb repulsion, effectively reduces the attractive bosonic glue by an unknown amount since the full theory including the retarded attraction at the Fermi level with a strong static repulsion overall is intractable.  But there are approximations leading to a phenomenological parameter $\mu^*$ which is considered to be $\mu^* \sim 0.2$ for normal metals~\cite{Allen1975,Allen1983}.  This $\mu^*$ effectively suppresses the attractive glue, and crudely speaking, one could assume that the $\lambda_b$ is reduced by $\mu^*$ to produce an effective lambda, given by $\lambda^* \sim \lambda_b - \mu^*$.  This, in fact, kind of explains why SC is absent in alkali metals such as Na and K or even Cu (where the phonon-induced $\lambda_b <0.2$, and hence the effective coupling is zero, producing no superconductivity).

Could the $\mu^*$ effect save the situation for the plasmon-induced metallic superconductivity and provide a meaningful answer within the nominal BCS heuristics just by replacing $\lambda_p$ with $\lambda_p-\mu^*$?  For $r_s \sim 6$ (which is typical for a 3D metal), assuming $\mu^*\sim 0.2$, we have $\lambda_p \sim 1.2$ and $\lambda_p^* \sim 0.2$, producing $T_{cp} \sim 10^4$~K$\cdot \exp(-5) \sim 70$~K, still far too high (although much less ridiculous than $1000$~K).  In addition, the basic theoretical conundrum is not resolved since for large enough $r_s$, the prediction would still be a high $T_c$, perhaps not $1000$~K, but $100$~K.  For example, an interaction-driven  3D Wigner crystal happening at $r_s \sim 100$ would be a superconductor with $T_{cp}\sim 10$~K.  In fact, such a crystalline superconductor would be a super-solid as it breaks both $U(1)$ and translational invariance.

We believe that arbitrarily increasing  $\mu^*$ by hand just in order to suppress plasmon-induced superconductivity at arbitrary $r_s$ is akin to saying that plasmons by itself  cannot induce superconductivity  in metals although the BCS theory (without any $\mu^*$ by definition) manifestly predicts a high-$T_c$ metallic superconductor.  The real problem, as discussed in the next section, is that there is an additional conceptual theoretical issue that all current theories of superconductivity are based on the Migdal theorem, which asserts that vertex corrections are negligible thus ruling out all high frequency contributions to SC pairing.  This theorem is valid for acoustic phonons where $\varepsilon_F\gg \omega_D$, but is simply invalid for plasmons since $\varepsilon_F\sim \omega_p$, and thus ``integrating out of high frequency contribution'' is neither allowed nor meaningful.  This is because the superconductivity is arising from electrons themselves, thus fundamentally ruling out a separation between the low and high frequency contributions, which is the key to the Migdal theorem enabling an (in principle) quantitative theory for superconductivity, which is the Migfal-Eliashberg theory to be discussed in the next section.

We conclude this section by asserting that although the naive BCS theory applied to the problem of the plasmon-induced metallic superconductivity predicts unreasonably high $T_c$,  there is no possible resolution of this conundrum within the BCS-Migdal-Eliashberg paradigm since the Hamiltonian allows no separation of energy scales, and as such vertex corrections and $\mu^*$ effects are leading order effects and cannot be neglected (or crudely approximated) as is done in all theories of superconductivity.

We discuss the Migdal-Eliashberg theory next for the plasmon-induced superconductivity elaborating on the findings of this section with more substantive technical details.

\section{Eliashberg-Migdal theory}\label{sec:eliashberg-migdal}
\subsection{Eliashberg framework for Coulomb-induced SC}
The Gorkov-Eliashberg formalism~\cite{carbotte1990properties} for calculating $T_c$ for electron-phonon superconductivity is one of the success stories of condensed matter physics. 
However, the role of Coulomb interactions in superconductivity~\cite{bogoliubov1960new} remained only partially understood even after the original BCS theory of superconductivity~\cite{schrieffer2018theory}, which was formulated for electrons in momentum space. The central puzzle was how the very weak phonon-mediated attraction can overcome the much stronger Coulomb repulsion~\cite{bogoliubov1960new}. This issue was resolved by considering the retardation in time of the phonon-mediated interaction~\cite{morel1962calculation} within the Gorkov-Eliashberg formalism~\cite{carbotte1990properties}. Intuitively, the idea is that the phonon-mediated attraction is long-ranged in time compared to the Coulomb repulsion, which is instantaneous. This allows electrons to form Cooper pairs by passing through a particular point in space at a later time to avoid the Coulomb interaction.
The intriguing consequence of this argument~\cite{morel1962calculation} is that Cooper pairing can occur even from an interaction remaining purely repulsive at all frequencies~\cite{morel1962calculation}. However, as apparent from the previous sections and will be discussed 
in more detail here, the role of Coulomb interactions continues to be poorly understood and calculationally intractable except using drastic and invalid approximations.

The Eliashberg theory mentioned above, which is required to properly account for Coulomb interactions, 
is characterized by the Eliashberg spectral function~\cite{carbotte1990properties} that is written as 
\begin{align}
\alpha^2F(\nu)=N(0)\int_{FS}dk dk'|g_{kk'}|^2(-\pi)^{-1}Im D(k-k';\nu),
\end{align}
where $D$ is the Boson propagator and $N(0)$ is the Fermi surface DOS and the integral averages over the Fermi surface.
As an aside, it should be noted that the formalism can be generlized to higher angular momentum channels (e.g.\ KL SC) but including the appropriate form factors in the integral for the spectral function. 
For the case of pairing by electron-phonon interactions, the above expression can be reformulated~\cite{kirzhnits1973description,allen1988total} in terms of the total screened Coulomb interaction $W$  according to the relation:
\begin{align}
\alpha^2F(\nu)=N(0)(-\pi)^{-1}\int_{FS}dk dk'Im[W(k-k';\nu)],  
\end{align}
where $W(k,\nu)=V_c(k)[1-\Pi(k;\nu)V_c(k)]^{-1}$ is the RPA screening of the Coulomb  interaction,  $V_c(k)$ is the bare Coulomb interaction and $\Pi(k;\nu)$ is the bare polarizability (which may or may not include phonons depending on the application of interest). The above pairing interaction has been used~\cite{rietschel1983role} to predict superconductivity in the electron gas from purely repulsive Coulomb interaction (i.e. without any phonons or other Bosons) treated within the RPA approximation and in a sense includes pairing mediated by both plasmons and electron-hole pairs. This was followed by work that augmented the RPA interaction with a Kukkonnen-Overhauser local field corrections~\cite{kukkonen1979electron} to reduce the $T_c$ of superconductivity in much of the parameter regime of the electron gas ~\cite{buche1990superconductivity,takada1993s,cai2022superconductivity} (We note, and as already discussed in Sec.~\ref{sec:coulomb}, we do not believe that using any local field corrections in the theory is an improvement from a fundamental conceptual perspective since such static local field corrections would violate Ward identities and are uncontrolled approximations.). As an aside, we note that the KL mechanism of superconductivity~\cite{Kohn1965} fits within the RPA framework as well, with the focus shifting to higher angular momentum channels while ignoring the retardation effects. The angular momentum form factors eliminate the repulsive part of the interaction for short-ranged bare repulsion, $V_c$, allowing the attractive part to dominate. This section will focus on reexamining the estimated superconductivity in the interacting three-dimensional electron gas.

The microscopic interaction $W$ that leads to pairing can be written in imaginary frequency in terms of the Eliashberg spectral function 
\begin{align}
    W(i\nu)=W(\infty)-\frac 2 {\pi^2 N(0)} \int_0^\infty d\nu' \alpha^2 F(\nu')\frac{\nu'}{\nu'^2+\nu^2},
\end{align}
where $\nu$ is the imaginary frequency and the interaction $W(i\nu)$ has been averaged over the FS in a similar way to the Eliashberg spectral function. Such an averaging over the Fermi surface is justified when the momentum space structure of the Cooper pair is known to be simple. We will elaborate on this further at a later stage. The first term in the above interaction is the Coulomb potential, $W(\infty)=V_c$, which is instantaneous and therefore frequency independent. The second term is generated either by some Boson or by screening from other electron-hole pairs. Since Eliashberg theory involves a non-perturbative summation of diagrams, a lynchpin of the theory is provided by Migdal's theorem~\cite{Migdal1958}, which shows that vertex corrections can be suppressed by separating out a low-frequency part from the interaction i.e. $W=W_<+W_>$ where 
\begin{align}
    W_<(i\nu)=-\frac 2 {\pi^2 N(0)} \int_0^{\omega_c} d\nu \alpha^2F(\nu')\frac{\nu'}{\nu'^2+\nu^2},
\end{align}
would include conventional electron-phonon interactions. 
Vertex corrections involving $W_<(i\nu)$ can be ignored~\cite{Migdal1958} if either one of $\omega_c$ or $W_<$ is small. Neither is true for the rest of the interaction 
\begin{align}
    W_>(i\nu)=V_c-\frac 2 {\pi^2 N(0)} \int_{\omega_c}^\infty d\nu' \alpha^2 F(\nu')\frac{\nu'}{\nu'^2+\nu^2},
\end{align}
which includes the bare Coulomb interaction.
The Eliashberg formalism~\cite{carbotte1990properties} for conventional superconductivity focuses on $W_<$ which is used to generate the so-called $\lambda=N(0)W_<(0)$, while the high-frequency part of the interaction $W_>$ is approximated by a Coulomb pseudopotential $\mu^*$~\cite{morel1962calculation},
whose value is believed to be $\mu^*\sim 0.15-0.2$ for most metals. This leads to a reasonably accurate prediction of $T_c$ for conventional superconductors when $\lambda\gtrsim 1$, which are the superconductors with respectable transition temperatures. However, this framework also assumes that the screening of the electron Boson interaction is well-approximated by RPA, which is not necessarily a bad approximation in many cases for nomal metals. It is clear from the literature of electron-phonon superconductivity that the Eliashberg formalism with RPA screened electron-phonon interactions~\cite{carbotte1990properties} with a Coulomb pseudopotential of $\mu^*$ and a low Debye frequency clearly is a successful theory of superconductivity. The cutoff $\omega_c\sim\omega_D$, which is the range of frequency over which the pairing operates in this formalism can be chosen to be much smaller than $E_F$ allowing Migdal's theorem to eliminate vertex corrections~\cite{Migdal1958}. In this work, we consider the effect of electron-electron interactions within RPA at larger $\omega_c$ to illustrate the role of vertex-corrections play at such $\omega_c$.  

The interpretation of $\omega_c$ i.e. the splitting $W=W_<+W_>$, becomes somewhat more complicated where the pairing arises from 
electron-electron interactions rather than an external Boson line. The Migdal prescription~\cite{Migdal1958} in the latter case can be viewed as splitting the interaction line in each diagram based on a frequency cutoff, which is not trivial to apply to electron-electron interactions. However, the renormalization group framework~\cite{shankar1994renormalization,metzner2012functional} maybe viewed as Fermions with a energy $|\omega|<\omega_c$ while  $W_>$ is the Fermion vertex consisting of all Fermion lines with $|\omega|>\omega_c$~\cite{metzner2012functional}. This interaction vertex is the building block of the Eliashberg equations~\cite{carbotte1990properties}. The interaction $W_<$ then necessarily contains at least one low-energy electron-hole pair in the Fermi shell. The superconducting transition temperature can be computed from the two-particle irreducible Cooper pair vertex, which be computed in principle irrespective of $\omega_c$, though such a calculation would not be subject to perturbation theory. On the other hand, Migdal's theorem~\cite{Migdal1958} tells us that the Eliashberg equations can be used to determine the two-particle irreducible vertex in the Cooper channel in the energy shell $|\omega|<\omega_c$. 

The choice of $\omega_c$ is a balancing act. Choosing $\omega_c$ too high introduces uncontrolled errors from ignored vertex corrections. A low choice of $\omega_c$, where Migdal's theorem~\cite{Migdal1958} still allows one to ignore verex corrections,  likely underestimates the transition temperature by restricting the choice of $\Delta$.   This is because, as will become clear in the next sub-section the Eliashberg equation for solving $\Delta(\nu)$ can be viewed as a variational minimization where reducing $\omega_c$ constrains the set of allowed SC states $\Delta(\nu)$. It is known for conventional SC~\cite{morel1962calculation} that the $T_c$ estimate for systems including both the screened Coulomb interaction and phonons, where the Coulomb pseudopotential $\mu^*$ is related to $\textrm{log}(E_F/\omega_D)$ and the Fermi energy $E_F$ served as the cutoff $\omega_c\sim E_F$, increases with decreasing $\omega_c$.  The numerical results for plasmon mediated SC in this work will show a similar trend of $T_c$ decreasing with decreasing $\omega_c$. Ultimately, the fact that we do not know $\omega_c$ leads to an uncertainty in $T_c$. However, this is not completely different from the uncertainty in $T_c$ from the value of the Coulomb psuedo-potential $\mu^*$~\cite{carbotte1990properties}, which is typically not known. The main reason the Coulomb pseudopotential formalism is successful is because it typically depends on the Fermi liquid properties of the normal state, which appears to not vary too strongly between systems. The introduction of $\omega_c$ is more microscopic than the Coulomb pseudopotential, in the sense that the logarithmic cut-off dependence~\cite{morel1962calculation} is built into it. There is more uncertainty in $\omega_c$, but in most cases the $T_c$ variation is less sensitive to $\omega_c$ compared to $\mu^*$. The approach in this work to deal with the uncertainty will be to compare the superconductivity in systems that can be expected to have similar electronic properties at high energy and choose the $\omega_c$ where one of the reference systems is not superconducting. This will still allow us to compare the extent to which different systems superconduct though the precise $T_c$ might be difficult to estimate.  

For simplicity, in this work, we will assume that $W_>$ calculated within RPA is a qualitatively correct approximation to the interaction vertex for $\omega_c$ small enough. With this approximation the one loop polarization diagrams in the Eliashberg equations modify $W_>\rightarrow W$ so that it is justified to use the full RPA interaction for $W$ with the external Fermions satisfying $|\omega|<\omega_c$.

\subsection{Variational solutions the Eliashberg equations}

 Ignoring quasiparticle renormalization (which is likely not critical for this case), the gap equation part of the Eliashberg equation~\cite{marsiglio2008electron} can be written as
\begin{align}
\phi_m=-\pi  T \sum_{m'}\frac{U(\omega_m-\omega_{m'})}{|\omega_{m'}|}\phi_{m'}.\label{eq:Eliash}
\end{align}
where $U$ in the above is the Fermi surface averaged screened Coulomb interaction vertex, which can potentially include vertex corrections~\cite{buche1990superconductivity,takada1993s,cai2022superconductivity}. Note that in order to be consistent with Migdal's theorem~\cite{Migdal1958}
as discussed in the last subsection, we will restrict the sum to $|\omega_m|<\omega_c$. Since superconducting $T_c$ can often be exponentially smaller than $\omega_c$, the above equation can be difficult to solve. On the other hand, it is possible to find bounds on $T_c$ by calculating appropriate integrals following the spirit of the McMillan equation~\cite{mcmillan1968transition}. To do this, we define $\psi_m=\phi_m/\sqrt{\omega_m}$, so that the gap equation can be written in symmetric form
\begin{align}
\psi_m=-\pi  T \sum_{m'}\frac{U(\omega_m-\omega_{m'})}{\sqrt{|\omega_{m'}||\omega_{m}|}}\psi_{m'}.\label{eq:Eliashsymm}
\end{align}
The above equation has a solution if the symmetric matrix on the RHS has a lowest eigenvalue below $-1$. This minimal eigenvalue of the matrix can be estimated variationally by minimizing the quadratic form  
\begin{align}
\pi  T\sum_{m,m'>0}\frac{U(\omega_m-\omega_{m'})+U(\omega_m+\omega_{m'})}{\sqrt{|\omega_{m'}||\omega_{m}|}}\psi_m\psi_{m'}<-1,\label{eq:var}
\end{align}
where $\sum_{m>0} \psi_m^2=1$. This variational form plays a crucial role in this section and is the rationale for being allowed to drop the momentum integrals in the sum. The variational principle allows us to restrict the possible forms of $\phi_m$ and estimate a superconducting $T_c$, which is a lower bound on the transition temperature. In our case we are restricting $\phi_m=0$ to be zero for $|\omega_m|>\omega_c$ as as well as to be momentum independent over a momentum shell of width $k_c$ i.e. $\phi_m(k)=\Theta(k_c-|k-k_F|)\phi_m$. This ansatz is included in the fact that $U(\omega_m)$ is proportional to the screened Coulomb interaction averaged over a shell of width $k_c$ around the Fermi surface.

Note that the Eliashberg theory for conventional superconductivity~\cite{marsiglio2008electron} usually chooses $k_c\sim 0$ because the averaged interaction in the usual electron-phonon case does not have a strong momentum dependence. Unfortunately, this cannot be used for the Coulomb interaction which has divergences at small momenta.
%This issue is exacerbated by our use of the $T=0$ RPA polarizability function because of which the smoothening of the singularity from thermal broadening is not accounted for in our results.
However, the cutoff $k_c$ is partly an artifact of the ansatz that is used in the variational solution and would not appear in a numerical momentum dependent solution of the Eliashberg equation. The variational nature of the above equations ensures that the solution obtained with a finite value of $k_c$ is a bound on the result of such a numerical calculation of $T_c$.
The main reason that $k_c$ needs to be discussed in detail is because it turns out that the $T_c$ for many of the Coulomnb interaction based systems considered here increases substantially as $k_c$ approaches $0$. This implies that the $T_c$ obtained by a momentum dependent solution of the Eliashberg equation would be similarly large. On the other hand, the SC gap around the Fermi surface that would be obtained in a $T=0$ solution can be expected to limit the effect of small momenta in the vicinity of the Fermi surface and regulate the divergence in the Coulomb interaction. This leads to the choice of $k_c\gtrsim \Delta/v_F$, where $\Delta$ is the superconducting gap. An alternative interpretation of $k_c$ would be the inverse coherence length or inverse size of the Cooper pair. 
This argument doesn't directly help in the solution of the linearized Eliashberg equation used to determine $T_c$ because $\Delta(T=T_c)=0$.  However, assuming $\Delta(T=0)\sim T_c$, leads to the  estimate of  $k_c\sim T_c/v_F$. While this appears to 
rely on an assumption of $\Delta\sim T_c$ where $\Delta$ refers to the $T=0$ SC gap, this is still a reasonable estimate of the $T_c$ since practically we would be interested in a temperature where $\Delta(T)$ is a comparable to the $T=0$ (i.e. maximum) SC gap. Furthermore, this estimate, is still a lower-bound on $T_c$ because the estimate $\Delta\sim T_c$ over-estimates the SC gap (relative to its value at $T\sim T_c$) and an over-estimate of $k_c$, which in turn under-estimates $T_c$. It should be noted that the $k_c$ cut-off is not fundamental and would not be necessary in a $T=0$ Eliashberg calculation. Also, the argument above suggests that such Coulomb interaction driven superconductivity might differ significantly from $\Delta(T=0)$. Both of these are numerically involved interesting questions, which are beyond the scope of the current analysis because as will be seen in the numerical plots, the sensitivity to $k_c$ would not change the order of magnitude of superconductivity in most cases.

The variational equation Eq.~\ref{eq:var} is still quite numerically intensive to solve when the transition temperature is small because it leads to a large density matrix. Therefore, motivated by early works in superconductivity~\cite{bogoliubov1960new,morel1962calculation,mcmillan1968transition} we assume a 2 valued matrix for $\psi_m$ i.e. $\phi_m=C [A+(B-A)\Theta(\omega_m-\omega_D)]\sqrt{2\pi T}$, where $C$ is a normalization condition determined by the equation 
\begin{align}
C^2(2\pi T)[A^2\sum_{|\omega_m|<\omega_D}\omega_m^{-1}+B^2\sum_{|\omega_m|>\omega_D}\omega_m^{-1}]=1,    
\end{align}
where $\omega_D<\omega_c$.
This ansatz was recently compared~\cite{chubukov2019implicit} to more modern renormalization group approaches.
This ansatz allows us to write Eq.~\ref{eq:var} as
\begin{align}
&(2\pi T)^2[A^2\sum_{|\omega,\omega'|<\omega_D}+B^2\sum_{|\omega,\omega'|>\omega_D}-2 A B \sum_{|\omega|>\omega_D>|\omega'|}]\nonumber\\
&\frac{U(\omega-\omega')+U(\omega+\omega')}{|\omega\omega'|}<-\frac{2}{C^2}.\label{eq:var1}    
\end{align}
The above equation can be re-written as 
\begin{align}
[A^2  (D_1+2 D_4)+B^2 (D_2+2 D_5) -2 A B D_3]< 0,\label{eq:var2}    
\end{align}
where $D_{j=1,\dots,5}$ are appropriate coefficients that will be defined later.
This condition can be satisfied if the coefficients satisfy the condition 
\begin{align}
D_0\equiv D_3^2-(D_1+2 D_4)(D_2+2 D_5)>0.   \label{eq:Disc} 
\end{align}
Comparing Eqns.~\ref{eq:var1} and ~\ref{eq:var2}, the coefficients in the above equation are found to be  
$D_1=\int_{\pi T}^{\omega_D}\frac{U(\omega-\omega')+U(\omega+\omega')}{|\omega\omega'|}$, $D_3=\int_{\pi T}^{\omega_D}d\omega'\int_{\omega_D}^{\omega_c}d\omega\frac{U(\omega-\omega')+U(\omega+\omega')}{|\omega\omega'|}$,  $D_2=\int_{\omega_D}\frac{U(\omega-\omega')+U(\omega+\omega')}{|\omega\omega'|}$, $D_4= \int_{\pi T}^{\omega_D}d\omega/\omega$, and $D_5= \int_{\omega_D}^{\omega_C} d\omega/\omega$, 
where at finite temperature the integrals represent Matsubara sums according to the convention $\int d\omega\sim (2\pi T)\sum_{m}$. For the purpose of establishing the existence of superconductivity, it will suffice to work with the $T\rightarrow 0$ limit, where the integrals can be taken literally. We will however verify our conclusions against taking the Matsubara sum directly as well. 

\subsection{Application to the three dimensional electron gas}
We will now apply the fairly general framework above to the three dimensional electron gas within RPA. The starting point for this is the Lindhard polarizability in imaginary frequency~\cite{takada1993s}  
\begin{align}
&\Pi(q,\omega)=-N(0)\nonumber\\
&\left[\frac 1 2+\frac{1-z^2+u^2}{8 z}\ln\frac{(1+z)^2+u^2}{(1-z^2)+u^2}-\frac u 2 \tan^{-1}\frac{2 u}{u^2+z^2-1}\right],
\end{align}
where $z=q/2 k_F$, $u=\omega/q v_F$ and $N(0)$ is the Fermi surface DOS. 
The RPA screened interaction 
\begin{align}
W(q,\omega)=[V_c(q)^{-1}-\Pi(q,\omega)]^{-1},    
\end{align}
where $V_c(q)=N(0)^{-1} (9\pi/4)^{-1/3} r_s/\pi z^2$ is the bare Coulomb interaction and $r_s$ is the electron gas parameter.
Scaling the above interaction by the density of states $N(0)$ to obtain the interaction $U$ which appears in the Eliashberg gap equation~\ref{eq:Eliash}, we can write $U(q,\omega)=N(0)W(q,\omega)$ as
\begin{align}
U(q,\omega)=U(q,\infty)[1-U(q,\infty)\varpi(q,\omega)]^{-1},\label{eq:Ufull}
\end{align}
where we have taken advantage of the fact that $U(q,\infty=N(0)V_c(q)$ is the scaled bare Coulomb interaction and 
\begin{multline}
\varpi(q,\omega)\equiv N(0)^{-1}\Pi(q,\omega)=-
\left[\frac 1 2\right.\\\left.+\frac{1-z^2+u^2}{8 z}\ln\frac{(1+z)^2+u^2}{(1-z^2)+u^2}-\frac u 2 \tan^{-1}\frac{2 u}{u^2+z^2-1}\right].    
\end{multline}
Integrating this over the momentum shell $k_c$ discussed in the previous sub-section leads to the frequency dependent interaction for Eq.~\ref{eq:Eliash}.
Since most expressions involve dimensionless parameters, it is useful to note that the frequency $\omega$ in the gap equation is in units where $E_F=v_F k_F/2=1/2$.

Interestingly, in the low frequency regime $\omega\ll v_F q$ that dominates superconductivity, the polarizability has a singular frequency dependence 
\begin{align}
\varpi(q,\omega)\approx -\frac 1 2-\frac {1-z^2} {4 z}\ln\frac{1+z}{1-z}+\frac{\pi |u|}{2}.    
\end{align}
The first part of the above expression is the familiar static Lindhard function. The next term is the lowest order imaginary frequency correction, which is suggestive of ohmic dissipation. Applying this to the scaled screened Coulomb interaction one gets 
\begin{align}
&U(q,\omega)\simeq [U(q,0)^{-1}-\pi |\omega|/2]^{-1}\nonumber\\
&\simeq U(q,0)+\pi U(q,0)^2|\omega|/2+\dots.\label{eq:Uapprox}
\end{align}
 Integrating over the momentum-shell leads to the approximation $U(\omega)=U_0+U_1|\omega|$. A numerical integration of the full screened Coulomb interaction Eq.~\ref{eq:Ufull} shows that this form is a good approximation for a large range of parameters.

\begin{figure}
    \includegraphics[width=\linewidth]{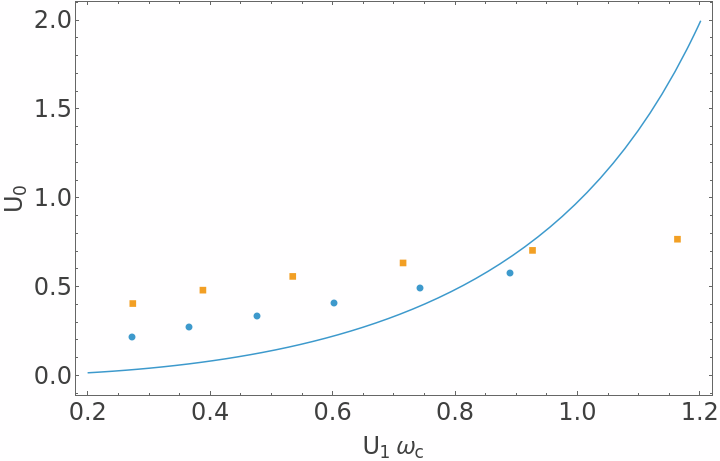}
    \caption{Dimensionless repulsive interaction $U_0$ versus the scaled dimensionless pairing interaction $U_1\omega_c$. The solid line represents the bound for the maximum $U_0$ that would support superconductivity. The circles and squares represent the numerically calculated values of $U_0$ and $U_1\omega_c$ for the 3D and 2D electron gases for $r_s=0.5,0.707,1.0,1.414,2,2.818$ respectively moving from left to right. The cutoff $\omega_c/E_F=1,0.5$ respectively for the 3D and 2D cases. The momentum shell width for the pair potential is assumed to be $k_c=10^{-3}k_F\sim T_c/v_F$. The electron gas is non-superconducting in this approximation at all the $r_s$ values except the last few. The value at $r_s=2.8$ superconducts assuming vertex corrections are negligible above $\omega_c\sim 0.8 E_F$ ($\omega_c\sim 0.3 E_F$ in 2D) but not for smaller $\omega_c$.}
    \label{fig:U0U1}
\end{figure}

We will now evaluate the transition temperature condition Eq.~\ref{eq:Disc} for the above interaction parametrized by $U_{0,1}$, which ultimately depend on the electron gas parameter. To evaluate Eq.~\ref{eq:Disc} in this case, we first substitute the linearized approximation for $U(\omega)$ from the previous paragraph into the integrals $D_{j=1,\dots 5}$ that appear in  Eq.~\ref{eq:Disc}. Since we are interested in whether such an interaction can support SC at any transition temperature, we take the $T\rightarrow 0$ limit and focus on the leading order i.e. $\left(\textrm{log}\frac{\omega_D}{\pi T_0}\right)^2$ term in Eq.~\ref{eq:Disc}. The resulting coefficient, after some algebra, can be written as: 
\begin{align}
& U_1 [2 U_0+U_1 (\omega_c-\omega_D)] (\omega_c-\omega_D)\nonumber\\
&-U_0 \ln\frac{\omega_c}{\omega_D}(1+2 U_1 \omega_D)>0.
\end{align}
Defining $y=\omega_D/\omega_c$ provides an upper bound on the repulsive interaction
\begin{align}
U_0<\textrm{max}_{y<1}\frac{U_1^2\omega_c^2(1-y)^2}{ -\ln{y} -2 U_1\omega_c(1-y+y \ln{y})},
\end{align}
that allows the existence of superconductivity. This is analogous to the fact that the Eliashberg coupling $\lambda$~\cite{mcmillan1968transition} must exceed the Coulomb pseudopotential $\mu^*$ to generate superconductivity. 
It should be noted that, the result of absence of superconductivity is technically based on a lower-bound on $T_c$, which leaves open the possibility of weak superconductivity with very low $T_c$. The parameters $U_{0,1}$ can be calculated for the electron gas using Eq.~\ref{eq:Ufull}. As shown in Fig.~\ref{fig:U0U1}, the electron gas can satisfy the above superconductivity condition for only one of the values of $r_s=2.8$ considered for $\omega_c=E_F$. Note that this SC for $r_s=2.8$ disappears for $\omega_c<0.9 E_F$, since reducing $\omega_c$ moves each point to lower $U_1\omega_c$.  One caveat for these results is that unlike the case of SC induced by electron-phonon coupling~\cite{marsiglio2008electron}, the electron gas points depend on the width $k_c=10^{-3}k_F$ of the momentum-space shell over which the order parameter $\phi_m(k)$ is uniform. In fact, smaller $k_c$ turns out to be more favorable to superconductivity so that for $k_c=10^{-2}k_F$ would render the range of $r_s$ non-superconducting while $k_c=10^{-6}k_F$ would lead to most of the range in Fig.~\ref{fig:U0U1} superconducting. However, this is an artifact of the momentum space singularities in the $T=0$ polarizability function $\varpi$ that would be rounded at finite temperature $T$. This limits the validity of our results to the regime  $k_c>T_c/v_F$. Therefore, for practical SC, one should limit to $k_c>10^{-4}k_F$. Even for parameters with superconductivity, the $T_c$ is expected to drop for small $\omega_c$, which can be checked by explicit numerical solution of the gap equation Eq.~\ref{eq:Eliash}. 
However, it  is clear from the results in Fig.~\ref{fig:U0U1} that $T_c$ is likely to be small or absent in much of the parameter regime of the 3D electron gas.

\subsection{Application to the two dimensional electron gas}\label{sec:eliashberg2D}
The analysis for the superconductivity in the two dimensional electron gas parallels the analysis in the previous subsection except that the dimensionless Lindhard polarizability is replaced by the corresponding function for the two dimensional electron gas~\cite{Stern1967}:
\begin{multline}
\varpi(q,\omega)=-(2 z)^{-1}[2z-f(z-iu)-f(z+iu)],    
\end{multline}
where $f(w)=w\sqrt{1-w^{-2}}$ and the square root is defined to choose the branch with the positive real part. The dimensionless Coulomb interaction in the two dimensional case is written as $U(q,\infty)=N(0)V_c(q)=r_s/z\sqrt{2}$, $r_s$ is the dimensionless 2D electron gas parameter defined in the last section. 

By repeating the calculations in the last sub-section with the above interaction and polarizability, we can calculate the critical value of $U_0$ as a function of $U_1\omega_c$ for the 2D electron gas at various $r_s$. The results of this calculation are demarcated by the square symbols in Fig.~\ref{fig:U0U1}. The 2D electron gas seems to be superconducting at a lower value of $r_s$ relative to the electron gas because of which we have plotted the results for $\omega_c=0.5 E_F$ instead of $\omega_c=E_F$ used for the 3D electron gas. One expects the relevant value of $\omega_c\ll E_F$ to be much smaller than $E_F$, so we do not expect the electron gas to be superconducting in this range of $r_s$ for either two dimensional or three dimensional electron gas. Also note that the cutoff factor $k_c$ introduced to account for the finite temperature effect on the polarization function $\varpi$ has a stronger $k_c/k_F$ correction in $2D$ as opposed to the $log(k_F/k_c)$ correction in 3D. This should be interpreted as the thermal corrections to RPA that are ignored in this work being a larger correction in 2D. 
\begin{figure*}
    
    \centering
    \includegraphics[width=0.22\linewidth]{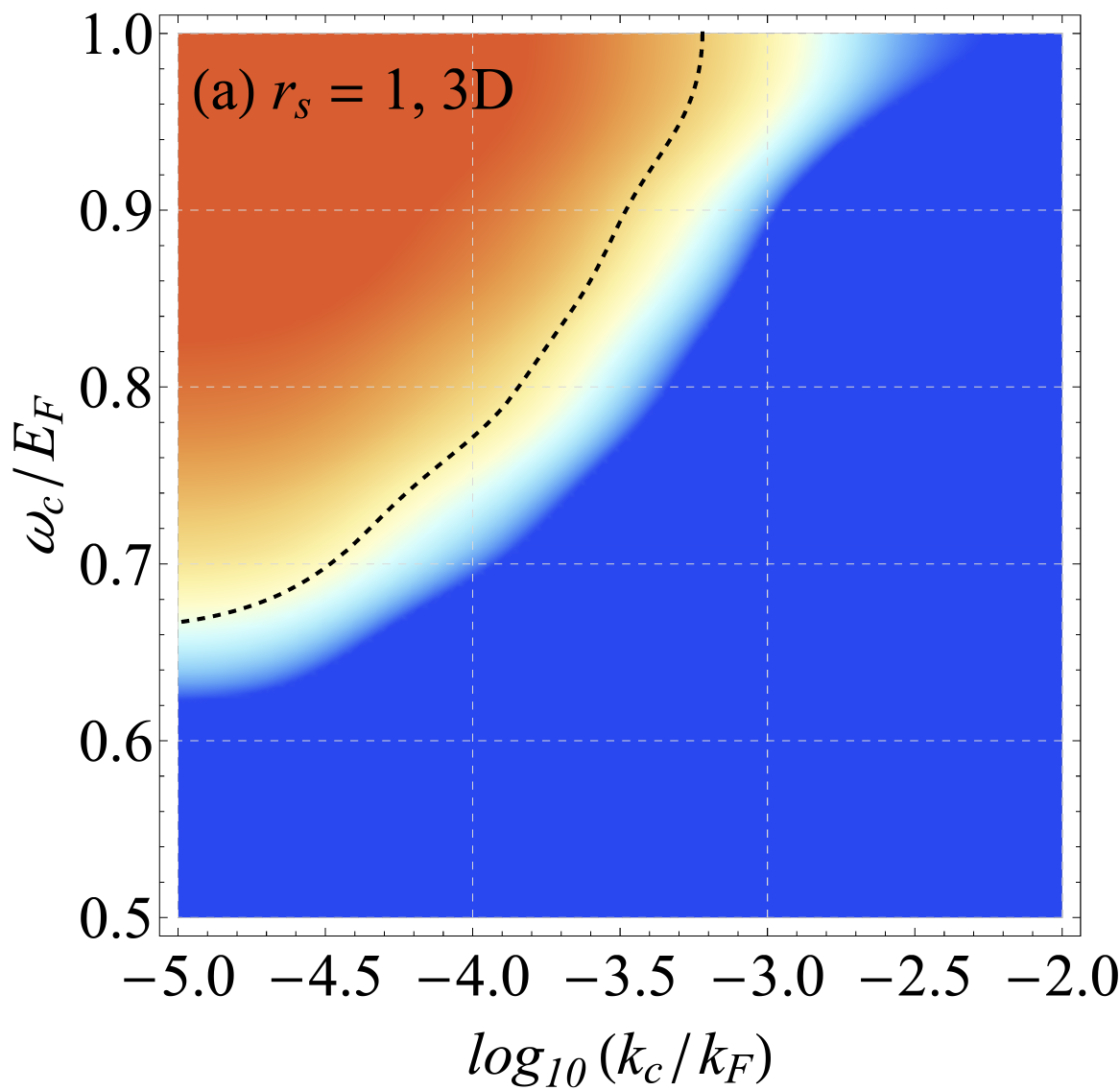}
    \includegraphics[width=0.22\textwidth]{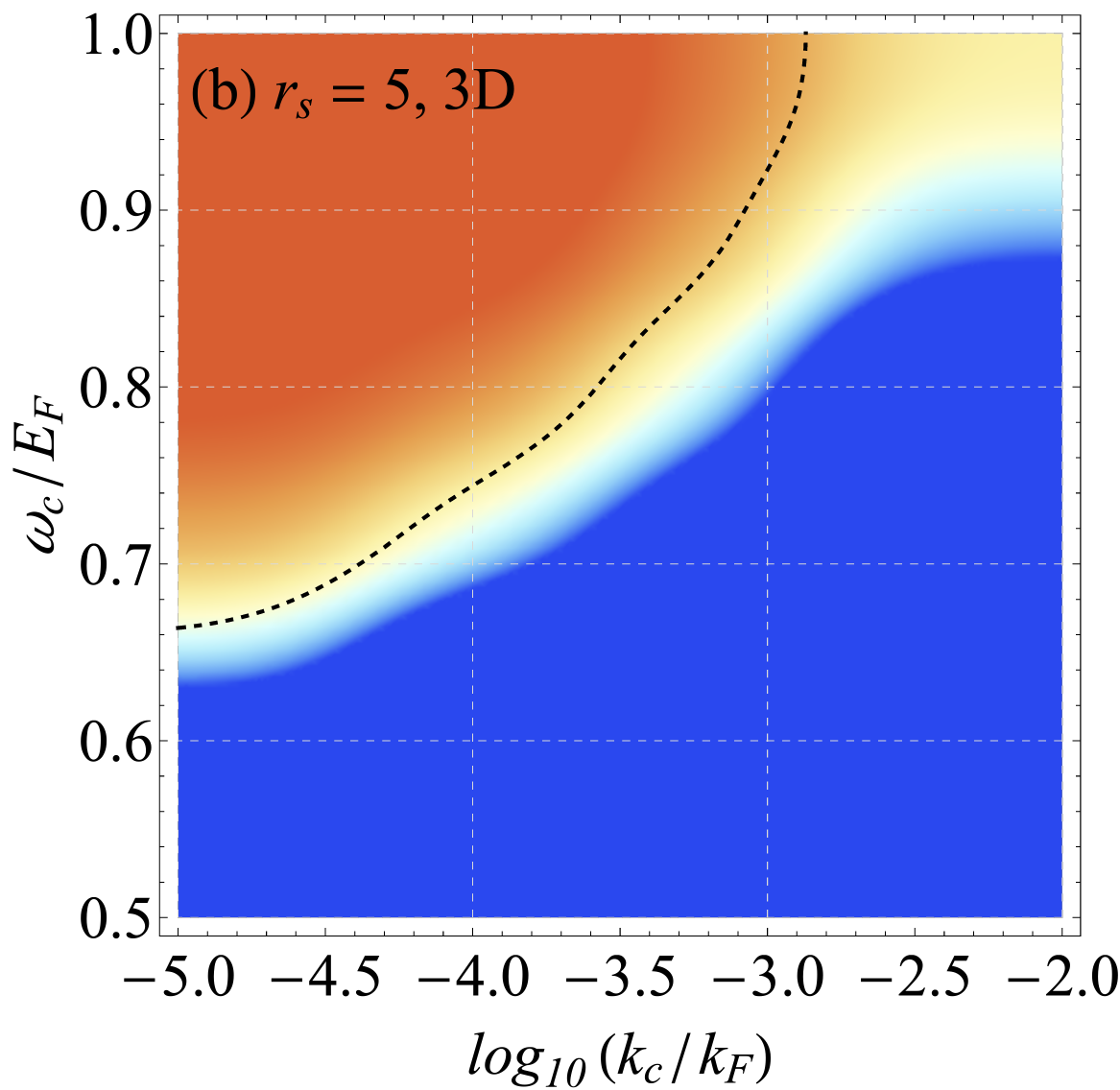}
    \includegraphics[width=0.22\textwidth]{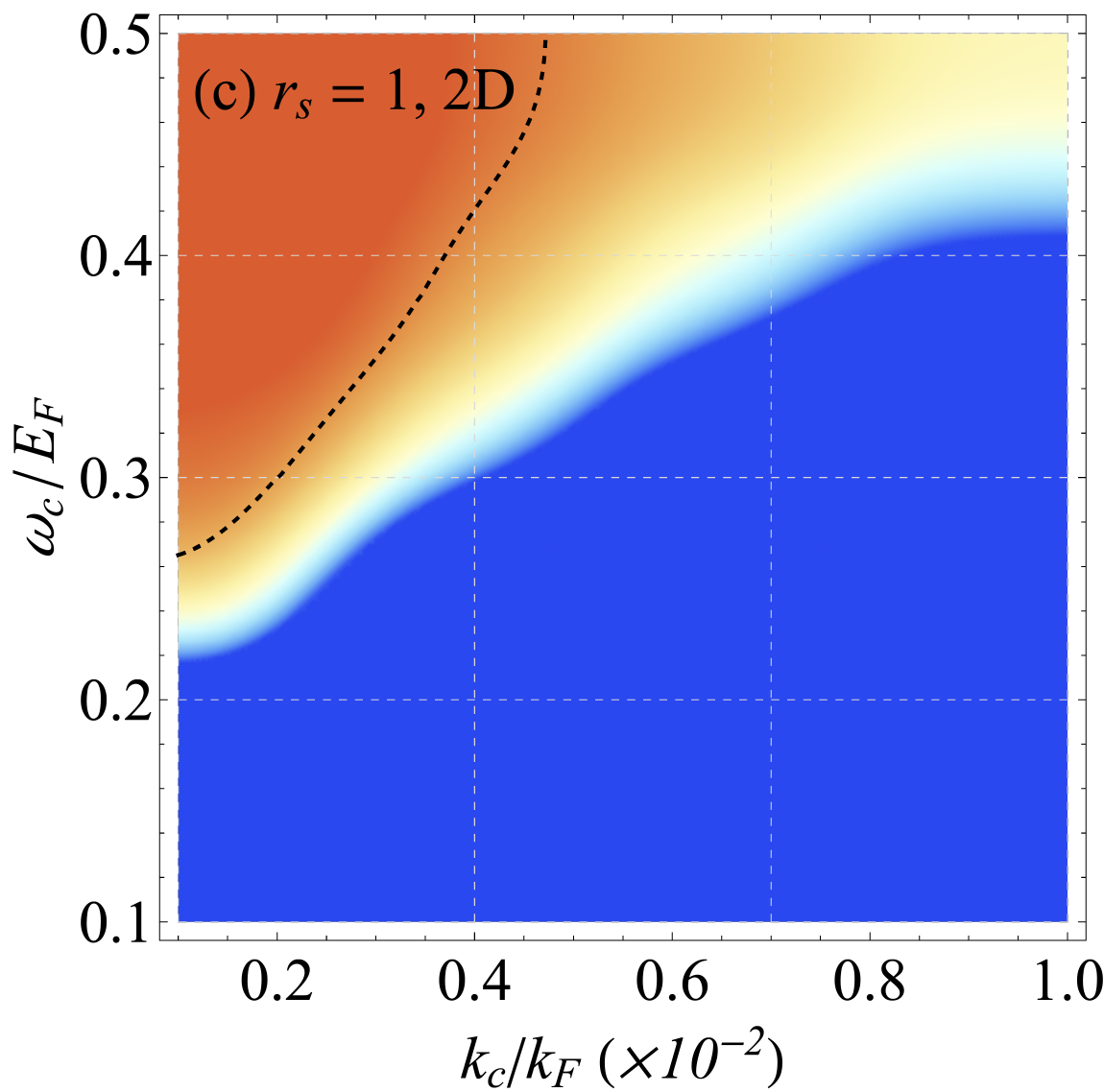}
    \hspace{0.3mm}
    \raisebox{-0.57mm}{\includegraphics[width=0.263\textwidth]{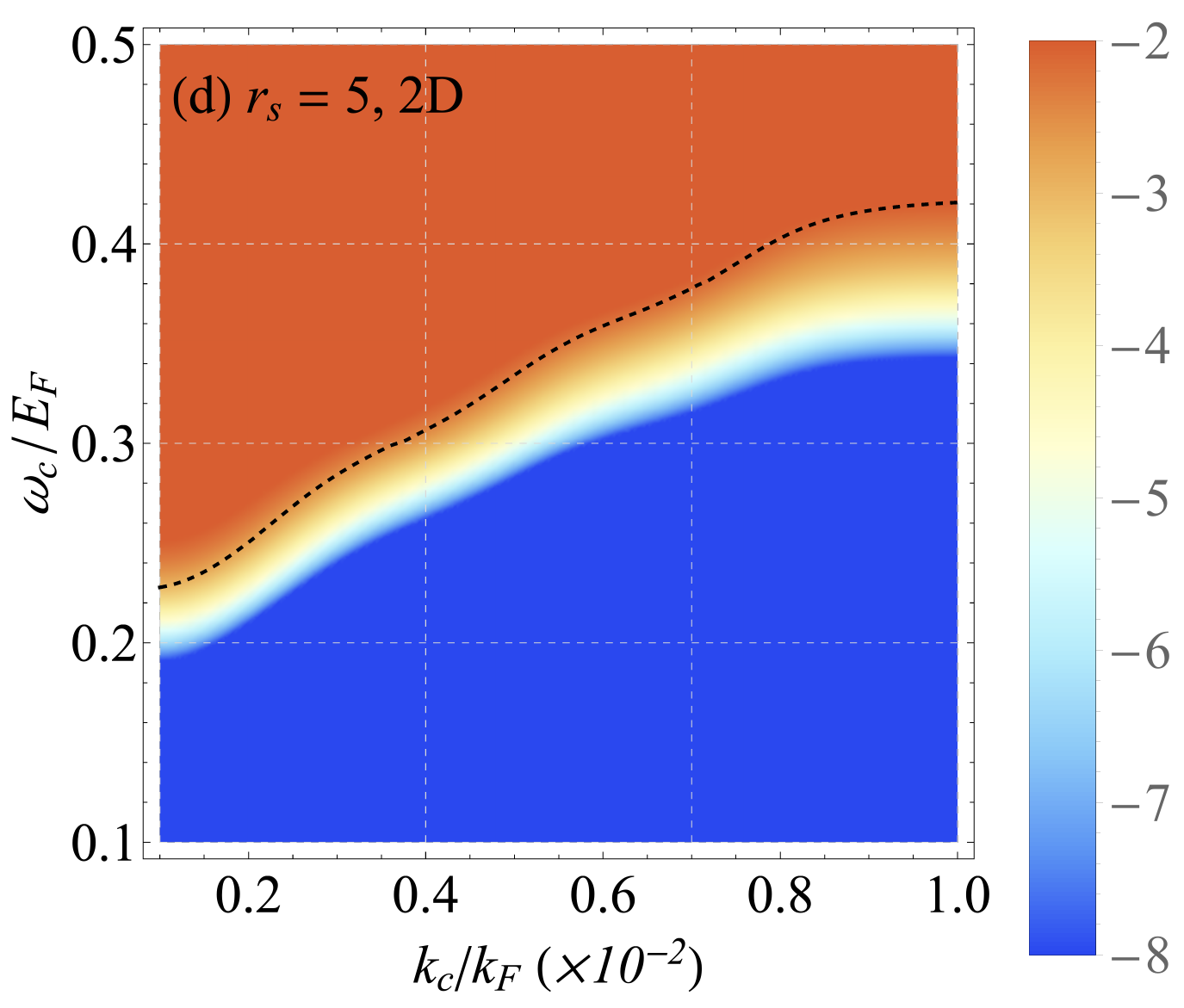}}
    \caption{Critical temperature $T_c$ as a function of the cutoff scales $k_c$ and $\omega_c$ for the electron gas with different parameters. The panels (a)-(d) respectively contain results for  (a) 3D electron gas at $r_s = 1$; (b) 3D electron gas at $r_s = 5$; (c) 2D electron gas at $r_s = 1$ and (d) 2D electron gas at $r_s = 5$. The color scale represents $\log_{10} (T_c/E_F)$, and the black line indicates the boundary $k_c = T_c/ v_F$, above which the results become unphysical. These results show the electron gas to be superconducting within RPA at a $T_c$ depending on $r_s$ and dimensions. However, the $T_c$ becomes negligible for $\omega_c\ll E_F$. As a reference for converting to $T_c$ to $K$ note that for an electron gas the Fermi energies in units K are $5.8\times 10^5, 2.3\times 10^4, 3.1\times 10^5$ and $1.3\times 10^5$ respectively from left to right.
    }
    \label{fig:Tccut}
\end{figure*}

\subsection{Results for $T_c$ in various regimes}\label{sec:eliashbergTc}
The previous subsections discussed conditions for the 2D and 3D electron gas to support a finite $T_c$ when the frequency dependence of the Fermi surface averaged screened Coulomb interaction $U(\omega)$ could be approximated to be linear (Eq.~\ref{eq:Uapprox}). In addition, the pairing potential $\phi_m$ was assumed to have a BCS-like ansatz. Although these approximations are reliable in regimes with low $T_c$, they are still uncontrolled approximations. The main motivation for these approximations is that the Eliashberg equations themselves, i.e. Eq.~\ref{eq:Eliash}, are notably difficult to solve when the transition temperature is small. Furthermore, these approximations make connections to the standard Allen-Dynes~\cite{Allen1975} and Coulomb pseudopotntial~\cite{morel1962calculation} formalism. 

In this sub-section, we present direct solutions to the Eliashberg equation Eq.~\ref{eq:Eliash}, where $U(\omega)$ is the Fermi-surface averaged screened Coulomb interaction for the 3D and 2D electron gases discussed in the previous two sub-sections. The transition temperature $T_c$ is determined by finding when a variational estimate of the lowest eigenvalue of Eq.~\ref{eq:Eliashsymm} drops below $-1$. Two complementary numerical approaches are employed depending on the relative scale of $T_c$ and the cutoff $\omega_c$.
In the regime $T_c \sim \omega_c$, the dimension of the matrix $U(\omega_m - \omega_{m'})$ is small, allowing the Eliashberg equation to be solved directly by computing its lowest eigenvalue. However, this regime typically fails to satisfy the physically relevant cutoff or limitation conditions $k_c \ll T_c / v_F$ and $\omega_c \ll E_F$, which will be discussed later, and therefore is not generally applicable.

In the regime $T_c \ll \omega_c$, the Eliashberg equation Eq.~\ref{eq:Eliashsymm} can be approximated by the integral eigenvalue problem
\begin{equation}
    \psi(\omega) = -\frac{1}{2} \int_{\pi T}^{\omega_c} d\omega' \,
    \frac{U(\omega - \omega') + U(\omega + \omega')}{\sqrt{\omega \omega'}} 
    \psi(\omega').
\end{equation}
Direct evaluation of this integral is numerically delicate due to singularities at $\omega = 0$ and $\omega' = 0$. To remove these singularities, we introduce the logarithmic change of variables $\omega = e^s$, which yields
\begin{equation}
    \tilde{\psi}(s) = -\frac{1}{2} 
    \int_{\ln(\pi T)}^{\ln(\omega_c)} 
    \big[U(e^s - e^{s'}) + U(e^s + e^{s'})\big] \tilde{\psi}(s') \, ds',
\end{equation}
where $\tilde{\psi}(s) = e^{s/2} \psi(e^s)$. This transformation renders the kernel smooth and non-singular, allowing for stable iterative numerical solution.

The resulting $T_c$ for the three dimensional electron gas at $r_s=1$ is shown in Fig.~\ref{fig:Tccut} (a). The $x$ and $y$ axes of the figure represent the two cut-offs  $\log_{10}(k_c/k_F)$ and $\omega_c/E_F$. As discussed at the beginning of this section, the momentum cut-off $k_c$ is a limit on the size of the superconducting gap. Fig.~\ref{fig:Tccut} (a) suggests that it is possible to increase $T_c$ at lower $\omega_c$ simply by going to smaller $k_c$, which is not a physical cut-off in the problem. However, this is actually a limitation of using the $T=0$ screened Coulomb interaction for our problem, which does not smear the momentum singularities. Therefore, the interaction enhancement below $k_c\gtrsim T_c/v_F$ are artifacts of the zero temperature approximation for the RPA screened Coulomb interaction. The other cut-off $\omega_c$, as also discussed in the beginning of a section is, unfortunately, a phenomenological parameter in the theory of SC and represents the unknown cut-off beyond which higher loop corrections must be included into Eliashberg theory. Since this is an unknown we plot the results in Fig.~\ref{fig:Tccut} (a) as a function of $\omega_c$. Consistent with the previous sub-sections, there is no significant superconductivity below $\omega_c<0.5 E_F$.
$T_c$ for the three-dimensional electron gas at $r_s = 5$, shown in Fig.\ref{fig:Tccut} (b) is qualitatively similar to the $r_s = 1$ case. The main difference between the two resuls are at large momentum cut-off $k_c$,  where $T_c$ is significantly enhanced compared to the $r_s = 1$ case. Thus, the enhancement of $T_c$ with increasing $r_s$ in 3D is mostly an artifact of an unphysical choice of $k_c$ related to the  zero-temperature approximation RPA screened Coulomb interaction.

The corresponding results for the 2D electron at $r_s = 1$ as shown in Fig.~\ref{fig:Tccut} (c), are notably different. Here, variations in $T_c$ have a linear dependence on $k_c$ as opposed to the logarithmic dependence in 3D. This motivates the use of a linear $k_c$ axis in panels (c,d) of Fig.~\ref{fig:Tccut}. When $\omega_c \sim E_F$ and $k_c$ is large, $T_c$ is significantly higher than in 3D. %; however, results for $k_c \lesssim T_c/v_F$ remain unphysical artifacts.
Similar to the 3D case (using the constraint $k_c \gtrsim T_c/v_F$ ), there is no significant superconductivity for $\omega_c \lesssim 0.1 E_F$ . For $r_s = 5$ in 2D, shown in Fig.\ref{fig:Tccut} (d), The variations relative to $r_s = 1$ are small: $T_c$ is slightly enhanced at large $k_c$ but decreases at small $k_c$.

\section{Summary and Conclusion}\label{sec:conclusion}

In this article, we have taken a deep dive into the question of electron-electron interaction induced conventional s-wave superconductivity in normal 3D metals using the simple jellium model in order to obtain generic conclusions not specific to any specific material band structure details.  We ignore phonons completely in the model, so any superconductivity arises entirely from the long range repulsive Coulomb electron-electron interactions.  We ignore the complications of higher orbital angular momentum `unconventional' superconductivity as happens for example from Friedel oscillations in high orbital angular momentum channels (not $s$-wave) in the so-called Kohn-Luttinger superconductivity which has exponentially low $T_c$.  Our interest is not the ground state pairing or the $T=0$ phase, but an experimentally relevant $T_c$ with laboratory consequences. Our conclusion using a number of complementary techniques is that such a regular conventional superconductivity is unlikely to occur in 3D metals although a direct uncritical application of the BCS theory would lead to unreasonably high $T_c$ ( $\sim 10^2$--$10^3$~K) in normal metals arising just from the electron-electron interaction, the so-called plasmon induced metallic superconductivity.
(Such uncritical predictions for plasmon-induced conventional metallic  superconductivity have indeed been made repeatedly in the literature---incorrectly in our opinion.)

Since the bare Coulomb interaction is by definition repulsive between electrons, the zeroth order fundamental question is whether electron-electron interaction can ever become attractive between electrons even in idealized well-controlled models, since without an effective attractive interaction, there is no pairing and no `regular and conventional' superconductivity.  This then brings up the important question of how the screened Coulomb interaction behaves in an electron gas, and whether it can ever become repulsive in some regions of the effective energy-momentum phase space.  The asymptotic behavior of quantum screening at $T=0$ is given by the zero-frequency Thomas-Fermi static screening, which goes as the following at long wavelength:
\begin{equation}\label{eq:epsilon-zeroomega}
    \varepsilon(q,\omega=0)=1+\frac{q_{TF}^2}{q^2},
\end{equation}
where $q_{TF}$ is the 3D Thomas-Fermi wavevector going as $q_{TF} \sim n^{1/6} \sim r_s^{-1/2}$, where $n$ is the 3D electron density and $r_s$ is the usual dimensionless coupling constant or the Wigner-Seitz radius.  For $q=0$, and high frequency, the dynamical screening goes as:
\begin{equation}\label{eq:epsilon-zeroq}
    \varepsilon(q=0,\omega)=1-\frac{\omega_p^2}{\omega^2},
\end{equation}
where $\omega_p$ is the plasma frequency.  We note that Eqs.~(\ref{eq:epsilon-zeroomega}) and (\ref{eq:epsilon-zeroq}) here, while being the leading order results in $q$ and $\omega$ respectively, refer to the limits $qv_F\gg \omega$ and $qv_F\ll \omega$, respectively.  We note that Eqs.~(\ref{eq:epsilon-zeroomega}) and (\ref{eq:epsilon-zeroq}) respectively indicate the low frequency screening and the high frequency anti-screening behaviors, with the screened interaction $u(q, \omega)= v_c (q)/ \epsilon (q, \omega)$ behaving as $u<v_c$ and $u>v$ for $\omega\ll qv_F$ and $\omega\gg qv_F$.  Note also that for $\omega\gg \omega_p$, the Coulomb interaction is unscreened.  All of these results are exact, and do not lead to any change in the sign of the effective or screened interaction, and this cannot lead to an effective attraction arising from electron-electron interaction.  We note the crucial feature that, although both the high- (Eq.~(\ref{eq:epsilon-zeroq})) and low- (Eq.~(\ref{eq:epsilon-zeroomega})) effective interaction remains repulsive, there is an intermediate frequency regime defined by $qv_F<\omega<\omega_p$, where the effective dynamically screened interaction becomes attractive since in this regime of $\omega<\omega_p$, the dielectric screening in Eq.~(\ref{eq:epsilon-zeroq}) is negative.  This emergent attraction between electrons is entirely a many-body retardation effect, which could, in principle, lead to pairing.
The actual attractive region, as described in Sec.~\ref{sec:coulomb}, is larger than the asymptotic considerations of Eqs.~(\ref{eq:epsilon-zeroomega}) and (\ref{eq:epsilon-zeroq}) indicate.
In 2D, the $q^2/q_{TF}^2$ factor in Eq.~(\ref{eq:epsilon-zeroomega}) is replaced by $q/q_{TF}$, but this does not change any of our considerations discussed above.

In Sec.~\ref{sec:coulomb}, we provided detailed numerical results for the dynamically screened effective interaction in several different widely used approximations, finding the somewhat surprising result that the effective interaction can indeed be attractive over a reasonable regime of momentum-energy phase space, thus allowing the minimal necessary condition for superconductivity.  In fact, our results show the dynamically screened Coulomb  interaction to be more persistently attractive than the simple considerations above suggest based on the asymptotic formula.  The best existing screening theory is RPA, which is exact in the high density (small $r_s$) limit, becoming progressively worse with increasing $r_s$.  We do provide some results with local field corrections, which are very similar to RPA, but local field corrections are uncontrolled and often arbitrary, thus most likely less trustworthy than RPA.  We find that the various simplifications of RPA, namely the plasmon-pole and hydrodynamic approximations, also give results very close to RPA  for the screened effective interaction.  Since the interaction develops an imaginary part (``damping'' of plasmons into electron-hole pairs) for larger momentum, only the part of the $(q, \omega)$ space where the imaginary interaction is zero and the real interaction attractive is of relevance to superconductivity in the BCS theory.  The existence of a finite regime in the $(q, \omega)$ space where $\operatorname{Re} u <0$ and $\operatorname{Im} u=0$ makes the discussion of interaction-induced (or equivalently plasmon-induced) superconductivity a meaningful exercise.

In Secs.~\ref{sec:bcs} and \ref{sec:eliashberg-migdal}, we provide the results  for $T_c$ based on the minimal BCS theory and the more complete Migdal-Eliashberg theory, respectively.  The naive BCS theory predicts absurdly high $T_c$ ($\sim 100$--$1000$~K) for normal metals arising from plasmon-induced pairing, with the fundamental reason behind this absurdity being the very high electronic energy scales of metals ($\sim 10^4$~K) and the basic electron-plasmon coupling strength, $r_s$, also being large, $r_s \sim 5$ for metals.  There are two related reasons for the manifest failure of the BCS theory here: the presence of the underlying omnipresent Coulomb repulsion (i.e.\ the so-called $\mu^*$ effect) suppressing $T_c$ and the vertex corrections.  Neither of these effects is included in the BCS theory, making its predictions unreliable.  An ad hoc inclusion of a $\mu^*$ could certainly push $T_c$ down, and in fact it is easy to suppress $T_c$ to zero simply by positing that $\mu^*$ is very large in this problem---e.g.\ a $\mu^* > 1$ will endure that normal metals do not undergo plasmon induced superconductivity, but a hypothetical system with very low density (and very large $r_s$) still might since such a large $r_s$ enables overcoming the repulsive $\mu^*$ effect.  This is, however, not doing theoretical physics, it is just data fitting, since there is no reliable method for calculating $\mu^*$ quantitatively.  The problem of vertex corrections is conceptually even greater.  The quantitative success of the theory of superconductivity for electron-phonon interaction is based entirely on the existence of a small parameter (even when the basic coupling is strong, as, e.g., for Pb where the dimensionless electron-phonon coupling $\sim 1.7$) which is essentially the ratio of the bosonic energy scale to the fermionic energy scale (or their group velocity ratio).  For phonons, this ratio is very small since the phonon Debye energy $\sim 10^2$~K and the electron Fermi energy $\sim 10^4$~K.  This is the celebrated Migdal theorem which ensures that vertex corrections are negligible, leading to the Migdal-Eliashberg theory.  There is no Migdal theorem whatsoever for the electron-electron interaction induced superconductivity by virtue of the fact that both the bosonic glue (e.g.\ plasmons) and the electrons, by definition, have the same energy scales, and hence the energy (or velocity) scale difference, which is the key to the Migdal theorem, does not exist for this problem by definition.  The nonexistence of the Migdal theorem makes any calculation of $T_c$ inherently unreliable for the current problem. In Sec.~\ref{sec:eliashberg-migdal}, we elaborate on this matter by providing a detailed analysis of the Migdal-Eliashberg theory as applied to 3D normal metals in the context of the electron interaction induced superconductivity. The application of Eliashberg theory, because of Migdal's theorem, requires us to identify a frequency cut-off $\omega_c$, where the Eliashberg gap equation is solved. Additionally, the $T=0$ approximation used for screening required us to consider a finite momentum space cutoff as well. For this analysis we effectively include both real and imaginary parts of the RPA dielectric function by using imaginary frequencies, so that pairing from both plasmons and electron-hole pairs are included.  Because the solution of the Eliashberg equation at very small $T_c$ can be subtle, we derive a condition for existence of superconductivity together with an estimate for $T_c$. Choosing reasonable values of the cut-off, $\omega_c$ we find that the electron-gas to be non-superconducting in much of the range of $r_s$. However, by calculating $T_c$ from numerically solving the Eliashberg equation in Sec.~\ref{sec:eliashbergTc} we find that a large cutoff $\omega_c\sim E_F$ can lead to an unrealistically large value of $T_c$. Corrections to RPA, which is the basis of our interaction model, are likely to become more important at higher $r_s$. The range of $r_s$ over which the gas is superconducting increases with lowering $k_c$. However, a $k_c$ much smaller than $10^{-3}k_F$ likely corresponds to a sub-kelvin gap superconductor. Furthermore, the superconducting gap or even the presence of superconductivity is found to be drastically reduced by lowering $\omega_c$, which is likely required by Migdall's theorem to avoid significant vertex correction effects. Ultimately, $\omega_c$ needs to be chosen to be small to avoid vertex corrections but too low a value of $\omega_c$ would under-estimate $T_c$.
%We find the basic low energy versus high energy separation, inherent in the Eliashberg theory, where the high energy processes are integrated out by virtue of the Migdal theorem, to be invalid for the current problem.  Because of such a lack of separation of energy scales, one could get any $T_c$ one wants if this separation is assumed arbitrarily, but such a result is unjustified theoretically.  
Our conclusion is that in all likelihood any plasmon induced $s$-wave metallic superconductivity has vanishing $T_c$. In fact, we find that even the $T=0$ system is not ordered for small eneough $\omega_c$, i.e., electron-electron interaction simply does not lead to $s$-wave superconductivity in 3D normal metals. It is still however possible that the approximations in our analysis of the Eliashberg equation miss a SC state with a very low $T_c$. We mention that our work does not rule out an exponentially low $T_c$ for normal metals in higher angular momentum channels arising from the Kohn-Luttinger mechanism, which we do not consider in the current work.

One could legitimately ask whether our conclusion of there being no plasmon-induced superconductivity applies to systems other than 3D jellium electron liquids, which is the explicit case we consider.  We believe that our work should apply to all situations where a jellium electron gas model applies since there is nothing specific to a 3D model of metals utilized in our work.  The fact that energy scales cannot be separated, and hence vertex corrections and Coulomb repulsion effects are important, but cannot be included in the BCS-Migdal-Eliashberg theories, applies to all situations, including much-studied 2D electron liquids, where the effective $r_s$ is often larger (and the effective $\omega_p$ smaller) because of 2D electron liquids, mostly being doped semiconductors, are inherently low-density metals.  The fact that 2D plasmon dispersion goes as $q^{1/2}$ and thus vanishes at long wavelength is a matter of profound inconsequence in our theoretical considerations since all the results involve some type of frequency-momentum integrations, and what matters is the plasmon energy at $k_F$ (or the plasmon energy at the momentum where it enters the electron-hole continuum).  In fact, the 2D theory for plasmon-induced metallic superconductivity is identical to our 3D theory, with exactly the same conclusions.  There have been claims~\cite{Takada1978} that having a bilayer 2D systems with acoustic plasmons~\cite{DasSarma1982} may facilitate plasmon-induced superconductivity, perhaps motivated by the fact that acoustic plasmons have the same linear-in-momentum energy dispersion as acoustic phonons.  But this claim is misguided and misleading because the Migdal theorem still does not apply, and the standard SC theories (i.e.\ BCS-Migdal-Eliashberg) would produce results very similar to what we get in our work, again with unreliable and unreasonably high $T_c$.  The problem of the inapplicability of the Migdal theorem does not become any less stringent just by virtue of having acoustic plasmons in the system since the bosons and the fermions are all arising from exactly the same Hamiltonian. As described in Sec.~\ref{sec:eliashberg2D}, the Eliashberg-Migdall analysis described in the previous paragraph can be extended to 2D electron gases as well. 
The results are consistent with the argument in this paragraph, despite the lower plasma frequency in 2D and the superconducting $T_c$ for the 2D electron gas is vanishingly small at $\omega_c\lesssim 0.1 E_F$. However, the values of cutoff where $T_c$ is suppressed needs to be chosen lower in the 2D electron gas compared to the 3D case.

The fundamental problem in the occurrence of plasmon-induced metallic superconductivity can be seen by doing a simple dimensional analysis.  The plasmon energy in both 2D and 3D metals goes as the square root of the electron density, $\omega_p\sim n^{1/2}$, and the Fermi energy $E_F$ goes as $k_F^2$, hence $E_F\sim n^{2/3}$ (in 3D), $n$ (in 2D).  So, the dimensionless ratio $\omega_p/E_F$ goes as $n^{-1/6}$ (in 3D) and $n^{-1/2}$ (in 2D).  Converting to the dimensionless coupling constant $r_s$ instead of density, we obtain $\omega_p/E_F \sim r_s ^{1/2}$ (3D); $r_s$ (2D).  This means that the Migdal condition of $\omega_p/E_F\ll 1$ is only achieved for $r_s\ll 1$ high-density limit.  But this is the precise limit when $T_c$ is exponentially small in the BCS theory (see Fig.~\ref{fig:Tcp}) because the $1/r_s$ factor in the exponential dominates in Eq.~(\ref{eq:Tcp-scaling}).  For $r_s\gg 1$, where the prefactor of Eq.~(\ref{eq:Tcp-scaling}) dominates and the exponential term is unimportant, the theory is completely uncontrolled by virtue of $\omega_p/E_F > 1$, implying all vertex corrections must be included in the theory.  The situation does not change at all if one considers a Dirac type linear dispersion where the coupling constant is often called $\alpha=e^2/v_F$, where $v_F$ is the  Fermi velocity of the Dirac electrons. (Note that $r_s$ for parabolic system is also $e^2/v_F$ basically, and $\alpha$ and $r_s$ denote the universal  dimensionless Coulomb coupling.)  Here a very flat band (e.g.\ moir\'e 2D layers) would have a very small effective $v_F$, and this very large $\alpha\gg 1$, again implying that $\omega_p/E_F\gg 1$ too, thus the theory is uncontrolled.  So, even leaving out the complications of the unknown repulsive effects of $\mu^*$, we have an inescapable;e conceptual theoretical problem in the sense that the limit ($r_s$, $\alpha \ll 1$) where the theory applies predicts exponentially low $T_c$, and for $r_s$, $\alpha\gg 1$ (i.e.\ flat bands or very large effective mass or very low carrier density), where the theory is meaningless, it predicts higher $T_c$.  The claims in the literature touting high $T_c$ generated by electron-plasmon interaction basically apply the theory uncritically in a regime where the theory is uncontrollably invalid because of the inapplicability of the Migdal theorem.

We mention that the manifest failure of the Migdal theorem for a pairing mechanism does not immediately lead to the conclusion that superconductivity may not occur due to that mechanism.  All it implies is the failure of our existing theories, the BCS-Migdal-Eliashberg theory, to describe any possible superconductivity arising from that mechanism.  Thus, the absurdity of blindly applying the BCS theory for plasmon induced pairing in 3D metals leading to a predicted $T_c \sim 10^2$--$10^3$~K only implies, by itself, that such a theory is not useful in discussing superconductivity in this context.  For example, if the problem could be solved exactly in some magical manner, where the issue of the inapplicability of the Migdal theorem is no longer relevant, we cannot rule out the existence of a plasmon-induced $T_c$.  The work in our Sec.~\ref{sec:eliashberg-migdal} indicates such a scenario to be extremely unlikely, but we have not solved the problem exactly.  We do believe that if the plasmon mechanism, in an unlikely (but not impossible) scenario, leads to superconductivity  in normal metals, the associated $T_c$ is likely to be impractically small by virtue of the fact that the direct effect of Coulomb repulsion will always be a huge challenge, i.e., the effective $\mu^*$ is likely to be large negating any attractive pairing trend induced by plasmons. The fact that the bare Coulomb interaction between electrons is repulsive added to the fact that plasmons are intrinsic electronic phenomena make it essentially impossible for the plasmon mechanism to lead to any `practical' conventional superconductivity in our opinion.

We emphasize that the full theory for plasmon-induced superconductivity in jellium metals must include not only vertex corrections but a complete treatment of the quasiparticle self-energy effects (going beyond the leading order approximations valid only for $r_s\ll 1$).  Such a theory does not exist, and is currently not feasible because it basically entails a full solution of the problem including all diagrams in the normal and anomalous channels.  Our work should be quantitatively calid for $r_s\ll 1$, where we find no superconductivity, and should be qualitatively and heuristically, but perhaps not quantitatively, in the strong coupling large *as well as $r_s\gg 1$) regime.

We now comment on the possibility of electron interaction induced superconductivity in strictly lattice strongly correlated Hamiltonians described by a tight binding Hamiltonian with mostly local interactions.  The paradigmatic model is the Hubbard model with onsite interaction and nearest neighbor hopping.  This model, which is complementary to our jellium based free electron Hamiltonian with long range Coulomb interaction, has been much studied in the literature motivated by high-$T_c$ cuprate superconductors.  Our work does not directly apply to this model although the inapplicability of the Migdal theorem as well as the $\mu^*$ problem exist here too.  The question of superconductivity in the repulsive Hubbard model is an extensively studied problem, which is well beyond the scope of the current work.  We mention two relevant works.  In Ref.~\cite{Raghu2010}, the authors conclude that for asymptotically weak interaction, where the weak coupling perturbative renormalization group should apply,  there is a $T_c$ which is exponentially small.  This is akin to the celebrated Kohn-Luttinger superconductivity in electron liquids.  The authors speculate that if their perturbative results can be uncritically extended to larger interaction strength, then a $T_c$ describing `regular' superconductivity (i.e.\ not exponentially small) may emerge.  But, this conclusion has been directly refuted in a later quantum Monte Carlo and DMRG calculations which assertively claims the absence of superconductivity in the repulsive Hubbard model for any interaction strength~\cite{Qin2020}.  But a later numerical work found that a slight modification of the tight binding kinetic energy term to include the next-nearest-neighbor hopping in a generalized Hubbard model does manifest superconductivity, but most likely with rather low $T_c$, and in higher angular momentum channel (e.g.\ $d$-wave)~\cite{Xu2024}.  We believe that the superconductivity discussed in the context of the Hubbard model in these (and possibly other) publications are essentially low-temperature superconductivity in higher orbital angular momentum, whose physics is connected to Kohn-Luttinger superconductivity in the electron liquid context, and not the `conventional' superconductivity with 'practical' transition temperature of interest in the current work.  Certainly, the superconductivity discussed in the strongly correlated lattice models of electron interactions is not plasmon-induced in the sense we describe in the current work, and is therefore outside the scope of the current work.  In this context, there are recent theories of superconductivity in various graphene and TMD based 2D layers with rather flatbands (which are very far from our electron liquid metals in the current work), where the superconductivity (with low $T_c$) may very well be arising from electron-electron interactions, but again such superconductivity is most likely a variation on the Kohn-Luttinger mechanism and is not generated by the attractive pairing glue of the virtual plasmon exchange between the electrons~\cite{Chou2021,Chou2025,Zhu2025}.

We conclude by asserting that the case for conventional superconductivity in metals with an experimentally observable $T_c$ arising from plasmon exchange (or equivalent mechanisms associated with electron-electron interactions)  has not been established at all in spite of there being many claims.
Our work shows that Coulomb coupling induced (or plasmon-induced) conventional $s$-wave superconductivity in normal metals to be unlikely.
Although we cannot decisively rule out the possibility of weak low-$T_c$ SC arising from Coulomb interactions (since the problem is not perturbatively amenable because of the inapplicability of the Migdal theorem), we can certainly rule out claims of plasmon-mediated SC with effectively high $T_c$, and our work establishes the strict guidelines that future theories of this problem must obey.

We do, however, add an important caveat here regarding the limitations of the theory.  When the Migdal theorem does not apply (i.e. the high frequency regime above the cut off in our Eliashberg equation in sec. IV), there is simply no controlled approximation for any rigorous conclusion.  While the Eliashberg theory is explicitly a strong coupling generalization of the simpler BCS weak-coupling theory, this is simply a statement on the coupling constant and not on the effect of vertex corrections which are still ignored in the theory.  We have sidestepped this issue by using the cutoff in the strong-coupling theory of sec IV and argued that the regime where the Eliashberg-Migdal theory actually applies (low cutoff values) $T_c$ is parametrically low for the plasmon-mediated superconductivity in the jellium model of metals.  We cannot however rule out that realistic band structure effects beyond the jellium model or other unknown physics not included in our theory cannot lead to a higher $T_c$ arising from the plasmon mechanism.  We speculate that this is unlikely to happen, but we cannot rule it out at this stage.  We do not consider possibilities od superconducting arising from spin fluctuations~\cite{chou2021correlation} or acoustic plasmons in multi-component systems~\cite{grankin2023interplay}, where the effective bosonic glue could in principle be much lower energy than the electronic Fermi energy, perhaps allowing the validity of the Migdal theorem.  We also do not consider the situation where both phonons and plasmons are present with the superconductivity arising mainly from the phonons, but plasmons contributing significantly to the physics~\cite{davydov2020ab}. Our work is entirely limited to a question of principle:  Can high-temperature s-wave superconductivity arise from the plasmons acting as the superconducting bosonic glue in jellium metals?  Our answer is a guarded 'no' with several caveats as we have noted above.

\section*{Acknowledgment}
The authors thankfully acknowledge helpful communications with Steve Kivelson and Shiwei Zhang.  This work is supported by the Laboratory for Physical Sciences. J.S. acknowledges support from the Joint Quantum Institute.

\appendix

\section{Details on the screened Coulomb interaction in 2D}\label{sec:coulomb-2D}

Similar to the 3D case considered in Sec.~\ref{sec:coulomb}, here we consider a two-dimensional electron gas with dispersion $\varepsilon_\mathbf{p}=p^2/2m$ and spin degeneracy 2.
At zero temperature, the Lindhard polarizability
\begin{equation}
    \chi(\mathbf{q},iq_n)=2\int \frac{d\mathbf{p}}{(2\pi)^2}\frac{f_\mathbf{p}-f_{\mathbf{p}+\mathbf{q}}}{iq_n+\varepsilon_\mathbf{p}-\varepsilon_{\mathbf{p}+\mathbf{q}}},
\end{equation}
can be calculated analytically ($f_\mathbf{p}=\theta(\varepsilon_\mathbf{p}-\mu)$).
Substitute $iq_n=\omega+i\epsilon,\epsilon\to 0$ for the retarded response and define dimensionless variables $\tilde q=q/2k_F$, $\tilde\omega=\omega/4\varepsilon_F$ and $\tilde\chi=\chi/d(\varepsilon_F)$ (where $d(\varepsilon_F)$ is the density of states at the Fermi energy $\varepsilon_F$, and $k_F$ is the corresponding Fermi momentum), we have~\cite{Stern1967,Ando1982}
\begin{multline}
    \tilde\chi(\tilde q,\tilde\omega+i\epsilon)=\frac{1}{2\tilde q}\Bigg[-i\sqrt{1-\left(\tilde q-\frac{\tilde\omega+i\epsilon}{\tilde q}\right)^2}\\+i\sqrt{1-\left(\tilde q+\frac{\tilde\omega+i\epsilon}{\tilde q}\right)^2}\Bigg]-1,
\end{multline}
where $\sqrt{1-(x\pm i\epsilon)^2}$ for real $x$ goes to $\sqrt{1-x^2}$ for $|x|<1$ and $\mp i\operatorname{sgn}(x)\sqrt{x^2-1}$ for $|x|>1$.

Under RPA, the dielectric function is approximated by
\begin{equation}\label{eq:epsilon-RPA-2D}
    \varepsilon_\text{RPA}(\mathbf{q},\omega)=1-v_c(q)\chi(\mathbf{q},\omega+i\epsilon)
\end{equation}
where $v_c(q)=2\pi e^2/q$ is the bare 2D Coulomb coupling.
The screened Coulomb interaction is
\begin{equation}
    u_\text{RPA}(\mathbf{q},\omega)=\frac{v_c(q)}{\varepsilon_\text{RPA}(\mathbf{q},\omega)}
\end{equation}
Defining the dimensionless interaction $u=\frac{\pi e^2}{k_F}\tilde u$, we can express everything using dimensionless parameters as
\begin{equation}
    \tilde u_\text{RPA}(\tilde q,\tilde\omega)=\frac{1}{\tilde q - \frac{r_s}{\sqrt{2}}\tilde\chi(\tilde q,\tilde\omega+i\epsilon)},
\end{equation}
where $r_s=(\pi a_0^2 n)^{-1/2}$ and $n$ is the 2D electron density.
The function is plotted in Fig.~\ref{fig:2D_RPA}.

For PPA we have~\cite{Stern1967,Ando1982}
\begin{equation}
    \begin{aligned}
        \varepsilon_\text{PPA}(\mathbf{q},\omega)&=1-\frac{1}{\omega^2}\left(\frac{2\pi ne^2 q}{m}+\frac{3}{4}q^2v_F^2\right)\\&=1-\frac{1}{\tilde\omega^2}\left(\frac{r_s}{\sqrt{8}}\tilde q+\frac{3}{4}\tilde q^2\right),
    \end{aligned}
\end{equation}
The resulting Coulomb interaction is plotted in Fig.~\ref{fig:2D_PPA}.

\bibliographystyle{plain}
\bibliography{references}

\end{document}